\documentclass[twocolumn]{svjour3}         
\smartqed  
\usepackage{graphicx}
\usepackage{amsmath}
\usepackage{bm}
\usepackage{amsfonts}
\usepackage{amssymb}
\begin{document}
\title{Breather-to-soliton and rogue wave-to-soliton transitions in a resonant erbium-doped fiber system with higher-order effects
}


\author{Lei Wang         \and
        Shen Li    \and Feng-Hua Qi}


\institute{Lei Wang$^*$ \at
               Department of Mathematics and Physics, North China Electric Power University, Beijing 102206, P.\ R.\ China \\
               $^*$Corresponding author; \email{50901924@ncepu.edu.cn}
           \and
           Shen Li \at
               School of Electrical and Electronic Engineering, North China Electric Power University, Beijing 102206, P.\ R.\ China
           \and
           Feng-Hua Qi \at
             School of Information, Beijing Wuzi University, Beijing 101149, P.\ R.\ China}

\date{Received: date / Accepted: date}

\maketitle
\begin{abstract}
Under investigation in this paper is the higher-order nonlinear Schr\"{o}dinger and Maxwell-Bloch (HNLS-MB) system which describes the wave propagation in an erbium-doped nonlinear fiber with higher-order effects including the fourth-order dispersion and quintic non-Kerr nonlinearity. The  breather and rogue wave (RW) solutions are shown that they can
be converted into various soliton solutions including the  multipeak soliton, periodic wave, antidark soliton, M-shaped soliton, and W-shaped soliton. In addition, under different values of higher-order effect, the locus of the eigenvalues
on the complex plane which converts breathers or RWs into solitons is calculated.
\PACS{42.65.Tg, 42.65.Sf, 05.45.Yv, 02.30.Ik}
\keywords{higher-order nonlinear Schr\"{o}dinger and Maxwell-Bloch system \and breather-soliton dynamics \and
rogue wave-soliton dynamics \and asymmetric rogue waves}

\end{abstract}

\section{Introduction}
\label{intro}\renewcommand{\theequation}{1.\arabic{equation}}\setcounter{equation}{0}
~~~~Optical solitons have attracted considerable interest of scientists around the world during the past decades~\cite{OP}. In optical fibers, there are two different types of solitons.
One is described by the nonlinear Schr\"{o}dinger (NLS) equation~\cite{NLSS,NLSS_1}, the mechanism of which is based on the balance between the second-order dispersion and Kerr nonlinearity. This type of soliton is called the NLS soliton. The other type of lossless pulse propagation  is the self-induced transparency (SIT) soliton in the two-level resonance medium, the dynamic response of which is governed by the Maxwell-Bloch (MB) system~\cite{MB}. When erbium  is doped with the core of the optical fibers,  the NLS soliton can exist with the SIT   soliton. The propagation of such pluse can be modeled by the NLS-MB system~\cite{NLSMB,NLSMB_1,NLSMB_2,NLSMB_3,NLSMB_4}. However, in an optical-fiber transmission system, one always needs to increase the intensity of the incident light field to produce ultrashort optical pulses~\cite{High,High_1,High_2,High_3,High_4}.  Thus, such higher-order effects as the higher-order dispersion, higher-order nonlinearity, self-steepening, and self-frequency shift, should be included~\cite{High,High_1,High_2,High_3,High_4}.

In this paper, we will work on  the higher-order NLS-MB system as follows~\cite{GNLS_MB,GNLS_MB1,GNLS_MB2,GNLS_MB3,GNLS_MB4},
\begin{equation} \label{GNLS_MB}
    \begin{aligned}
E_{z}=&i \,(E_{tt}+2\,|E|^{2}\,E)+i\, \tau\,(E_{tttt}+8\,|E|^{2}\,E_{tt}\\
    &+2\,E^{2}\, E_{tt}^{\ast}+6\,E^{\ast}\,E_{t}^{2}+4\,|E_{t}|^{2}\,E\\
    &+6\,|E|^{4}\,E)+2\,p\,,\\
    p_{t}=&2i\, \omega p+ 2\,E \eta,\\
   \eta_{t}=&-(E p^{*}+E^{*} p)\,.
       \end{aligned}
 \end{equation}
where the subscripts $z$, $t$ are partial derivatives with respect to the distance and time, the
asterisk symbol stands for the complex conjugate. $E$ denotes the normalized slowly varying amplitude
of the complex field envelope, $p=v_{1}v^{*}_{2}$  represents as the polarization, and $\eta=|v_{1}|^{2}-|v_{2}|^{2}$ means
the population inversion with $v_{1}$ and $v_{2}$ being the wave functions of the two energy
levels of the resonant atoms, $\omega$ is the frequency, $\tau$ is a small dimensionless real parameter.

Various solutions of System~(\ref{GNLS_MB}) have been discussed. For example,
based on the Lax pair  and Darboux transformation (DT), the multi-soliton solutions of System~(\ref{GNLS_MB}) have been presented in Ref.~\cite{GNLS_MB}.  The breather,  rogue wave (RW) and hybrid solutions of  System~(\ref{GNLS_MB}) have been obtained by means of the generalized DT  in Refs.~\cite{GNLS_MB1,GNLS_MB2,GNLS_MB3,GNLS_MB4}.  Solitons, breathers, and RWs are different types of
nonlinear localized waves and are central objects in diverse nonlinear
physical systems~\cite{ND,ND1,ND2,ND3,ND4,ND5,ND6,ND7,ND8,ND9,ND10}. The hybrid solutions, which are expressed in forms of the mixed rational-exponential functions,  describe the nonlinear superposition of the RW and breather~\cite{hi}. Recently, Liu \emph{et al.} have shown that the breathers can be converted into different types of nonlinear waves in the coupled NLS-MB system, including the multipeak soliton, periodic wave, antidark soliton, and W-shaped soliton~\cite{Liu}. These conversions occur under a  special condition in which the  soliton and a periodic wave in the breather have the same velocity~\cite{Liu}. In addition, the similar transitions have also been reported in some higher-order nonlinear equation of evolutions. Akhmediev \emph{et al.} have found that the breather solutions of the Hirota~\cite{A-Hirota} equation and fifth-order NLS~\cite{A-FNLS} equation  can be converted into soliton solutions  on the constant background, which does not exist in the standard NLS equation. Liu \emph{et al.} have revealed that the transition between the soliton and RW  occurs as a result of the attenuation of MI growth rate to vanishing in the zerofrequency perturbation region~\cite{Hirota,Hirota_1}.

It is natural to ask: Can the transition between the breather (or RW) and soliton occur in the higher-order coupled system, e.g., System~(\ref{GNLS_MB})? In this paper, we show that System~(\ref{GNLS_MB}) does have such transition when the  eigenvalues are located at the special locus on the complex plane. Further, we demonstrate that the breather can be converted into the multipeak soliton, periodic wave, antidark soliton, and M-shaped soliton while the RW can be transformed into the W-shaped soliton.

The outline of this paper will be as follows:   The  breather-to-soliton conversion and several types of transformed nonlinear waves will be studied in Section~2.  The RW-to-soliton conversion  will  be discussed in Section~3. Finally, our
conclusions will be addressed in Section 4.
\vspace{0cm}
\section{  Breather-to-soliton conversions } \label{sec:1}
\renewcommand{\theequation}{2.\arabic{equation}}\setcounter{equation}{0}
~~~In this section, we present different types of nonlinear wave solutions on constant background for System~(\ref{GNLS_MB}).
We omit the discussions of the components $p$ and $\eta$  because the types of them have the similar characteristics as $E$.
Based on the lax pair and DT~\cite{GNLS_MB,GNLS_MB1,GNLS_MB2,GNLS_MB3,GNLS_MB4} of System~(\ref{GNLS_MB}), the first-order symmetric breather solution reads as\\
\vspace{-0.4cm}
\begin{equation}\label{solution}
\begin{aligned}
E_{B}^{[1]}=&d\,(1+8\,n\,\frac{G_{B}^{[1]}+i\,H_{B}^{[1]}}{D_{B}^{[1]}})\,e^{i\,\rho}\,,\\
p_{B}^{[1]}=&\frac{1}{2} \,((E_{B}^{[1]})_{z}-i \,((E_{B}^{[1]})_{tt}+2\,|E_{B}^{[1]}|^{2}\,E_{B}^{[1]})-i\, \tau((E_{B}^{[1]})_{tttt}\\
&+8\,|E_{B}^{[1]}|^{2}\,(E_{B}^{[1]})_{tt}+2\,(E_{B}^{[1]})^{2}(E_{B}^{[1]})_{tt}^{\ast}+6\,(E_{B}^{[1]})^{\ast}\\
&(E_{B}^{[1]})_{t}^{2}+4\,|(E_{B}^{[1]})_{t}|^{2}\,E_{B}^{[1]}+6\,|E_{B}^{[1]}|^{4}\,E_{B}^{[1]}))\,,\\
\eta_{B}^{[1]}=&\frac{-2\,i\,\omega\,(p_{B}^{[1]})^{[1]}+(p_{B}^{[1]})_{t}^{[1]}  }{2\,E_{B}^{[1]}}\,.
\end{aligned}
\end{equation}
where\\
\vspace{-0.2cm}
\[\begin{aligned}
&G_{B}^{[1]}=2\,k_{1}\,k_{2}\,n\,\cos(t\,h_{R}+z\,\varpi_{B})
+2\,n\,\cosh(t\,h_{I}+z\,\varpi_{A})\\
&~~~~~~~~-k_{1}\,k_{2}\,h_{R}\,\sin(t\,h_{R}+z\,\varpi_{B})+h_{I}\,\sinh(t\,h_{I}+z\,\varpi_{A})\,,\\
&H_{B}^{[1]}=k_{1}\,k_{2}\,(b+2\,m)\,\cos(t\,h_{R}+z\,\varpi_{B})+(b+2\,m)\,\cosh(t\,h_{I}\\
&~~~~~~~~+z\,\varpi_{A})+k_{1}\,k_{2}\,h_{I}\,\sin(t\,h_{R}+z\,\varpi_{B})+h_{R}\,\sinh(t\,h_{I}\\
&~~~~~~~~+z\,\varpi_{A})\,,\\
&D_{B}^{[1]}=k_{1}\,k_{2}\,(|h|^{2}-\chi)\,\cos(t\,h_{R}+z\,\varpi_{B})
-(|h|^{2}+\chi)\,\cosh(t\,h_{I}\\
&~~~~~~~~+z\,\varpi_{A})-k_{1}\,k_{2}\,\Gamma_{A}\,\sin(t\,h_{R}+z\,\varpi_{B})-\Gamma_{B}\,\sinh(t\,h_{I}\\
&~~~~~~~~+z\,\varpi_{A})\,,
\end{aligned}
\]with\vspace{-0.2cm}
\[\begin{aligned}
&\rho=\,a\,z+b\,t\,,~~\lambda=m+i\,n\,,\\
&a=(b^{4}-12\,b^{2}\,d^{2}+6\,d^{4})\,\tau-(b^{2}-2\,d^{2})+\frac{4}{2\,\omega-b}\,,\\
&\chi=4\,m^{2}+4\,n^{2}+4\,m\,b+b^{2}+4\,d^{2}\,\,,\\
&\Gamma_{A}=2\,b\,h_{I}+4\,m\,h_{I}-4\,n\,h_{R}\,,\\
&\Gamma_{B}=2\,b\,h_{R}+4\,m\,h_{R}+4\,n\,h_{I}\,,\\
&\varpi_{A}=2(\kappa_{R}\,h_{I}+\kappa_{I}\,h_{R})\,,\\
&\varpi_{B}=2(\kappa_{R}\,h_{R}-\kappa_{I}\,h_{I})\,,\\
&\kappa=\frac{1}{2}(-b+2\,\lambda+(b^{3}-6\,b\,d^{2}+(4\,d^{2}-2\,b^{2})\,\lambda+4\,b\,\lambda^{2}\\
&~~~~~-8\,\lambda^{3})\,\tau)-\frac{1}{(b-2\,\omega)(\lambda+\omega)}=\kappa_{R}+i\,\kappa_{I}\,,\\
\end{aligned}
\]
\[\begin{aligned}
&h=2\,\sqrt{d^{2}+(\lambda+\frac{b}{2})^{2}}=h_{R}+i\,h_{I}\,.
\end{aligned}\]

Solution~(\ref{solution})  includes the hyperbolic functions $\sinh $\\
$(t\,h_{I}+z\,\varpi_{A})$ ( or $\cosh (t\,h_{I}+z\,\varpi_{A})$) and the trigonometric functions $\sin (t\,h_{R}+z\,\varpi_{B})$ (or $\cos (t\,h_{R}+z\,\varpi_{B})$), where $\kappa_{R}+\frac{\kappa_{I}h_{R}}{h_{I}}$  and  $\kappa_{R}-\frac{\kappa_{I}h_{I}}{h_{R}}$ are the corresponding
velocities. In this case, the hyperbolic functions and trigonometric
functions, respectively, characterize the localization and the periodicity of the transverse distribution $t$ of those waves. The
nonlinear structure described by  Solution~(\ref{solution}) can be seen as a nonlinear combination  of a soliton with the
velocity $\kappa_{R}+\frac{\kappa_{I}h_{R}}{h_{I}}$ and a periodic wave  with the  velocity $\kappa_{R}-\frac{\kappa_{I}h_{I}}{h_{R}}$.
Next, we will display various nonlinear wave structures   depending
on the values of velocity difference, namely, $\frac{\kappa_{I}(h_{R}^{2}+h_{I}^{2})}{h_{R}h_{I}}$.

If the velocity difference is not equal to zero, i.e.,  $\kappa_{I}(\frac{h_{R}^{2}+h_{I}^{2}}{h_{R}h_{I}})\neq0$ (or $k_{I}\neq0$),  Solution~(\ref{solution}) characterizes the
localized waves with breathing behavior on a plane-wave background (i.e., the breathers and RWs).  Further, if $m=-\frac{b}{2}$, we have the Akhmediev breathers
with $|n|<|d|$, the Kuanetsov-Ma solitons with $|n|>|d|$,
and the Peregrine soliton with $|n|=|d|$. Such solutions have been derived in
Refs.~\cite{GNLS_MB1,GNLS_MB2,GNLS_MB3,GNLS_MB4} (also see \textbf{Fig.~1}).

Instead, if  $\kappa_{I}=0$,  the wave described by  Solution~(\ref{solution}) is  composed of
a soliton and a periodic wave, where each has the same velocity $\kappa_{R}$. It should be noted that  the case $\kappa_{I}=0$ is equivalent to
 \begin{equation}\label{KZFC1}
    \begin{aligned}
&\frac{\varpi_{A}}{h_{I}}=\frac{\varpi_{B}}{h_{R}}\,,
  \end{aligned}
 \end{equation}
 i.e.,
\begin{equation}\label{situation11}
    \begin{aligned}
&\tau=\frac{\Theta_{1}}{\Theta_{2}}\,,
    \end{aligned}
 \end{equation}
where
\[ \begin{aligned}
&\Theta_{1}= (b-2 \omega ) \left((m+\omega )^2+n^2\right)+1\,,\\
    &\Theta_{2}=(b-2 \omega ) \left((m+\omega )^2+n^2\right)(b^2-4 b m-2 (d^2-\\
    &~~~~~~~~6 m^2+2 n^2))\,.
 \end{aligned}
\]

Eq.~(\ref{KZFC1}) [i.e., Eq.~(\ref{situation11})] implies the extrema of trigo-nometric and hyperbolic functions in
Solution~(\ref{solution}) is located along the same straight lines in the $(z,t)$-plane, which leads to the transformation of the breather into a continuous
soliton. Choosing different values of $\tau$ in Eq.~(\ref{situation11}), we display  the locus of the  real and imaginary parts of the eigenvalues on the complex plane in \textbf{Fig.~2}. It is found that the higher-order effect $\tau$ can alter the shape of the locus. By decreasing the value of $\tau$, we find that the quantity of branches of the locus  is reduced from three to two.

When the five parameters $m$, $n$, $\omega$, $\tau$, $b$ and $d$ satisfy Eq.~(\ref{situation11}), Expressions~(\ref{solution}) describe the soliton on constant background, as depicted in \textbf{Fig.~3}. This type of multi-peak structure is because of the mixture of a soliton and a periodic wave. The similar multi-peak structures have been observed analytically in the NLS-MB equations~\cite{Liu} and numerically in the AC-driven damped NLS system~\cite{mp,mp_1}. Further, to exhibit the effect of the real part ($m$) of the eigenvalue on the peak number, we plot \textbf{Fig.~4} where the soliton has  significantly less peaks than the one in Fig.~3. On the other hand, due to the different choices of the value of the  eigenvalue, \textbf{Fig.~5} shows the soliton with two main peaks, which is referred to as the M-shaped soliton in this paper. Note that the multi-peak solitons in Figs.~3$\sim$5 are the localized structures  along the $t$ axis. However, changing the  values of  the real and imaginary parts of eigenvalue, we display another type of solution that shows a type of nonlocal structure, i.e.,  the periodic wave (see \textbf{Fig.~6}). Therefore, it is concluded that the structure (peak number and  localization) of the multi-peak soliton can be controlled by
 the real and imaginary parts of eigenvalue, namely, $m$ and $n$.

In order to better understand this multi-peak structure of System~(\ref{GNLS_MB}), we will
extract separately the soliton and periodic wave from the mixed solution~(\ref{solution}). Specifically, the soliton exists  in isolation when $h_{R}$
vanishes, while the periodic wave  independently exists when $h_{I}$ vanishes. Correspondingly,
the  analytic  expressions read, for the  antidark soliton,
\begin{equation}\label{GNLShr}
\begin{aligned}
& E_{S}^{[1]}=&d\,(1+8\,n\,\frac{G_{S}^{[1]}+i\,H_{S}^{[1]}}{D_{S}^{[1]}})\,e^{i\,\rho}
\end{aligned}
\end{equation}
with
\[
\begin{aligned}
&G_{S}^{[1]}=2\,n\,\cos(2\,z\,\kappa_{I}\,h_{I})
+2\,n\,\cosh(t\,h_{I}+2\,z\,\kappa_{R}\,h_{I})\\
&~~~~~~~~+h_{I}\,\sinh(t\,h_{I}+2\,z\,\kappa_{R}\,h_{I})\,,\\
&H_{S}^{[1]}=\,(b+2\,m)\,\cos(2\,z\,\kappa_{I}\,h_{I})+(b+2\,m)\,\cosh(t\,h_{I}\\
&~~~~~~~~+2\,z\,\kappa_{R}\,h_{I})-\,h_{I}\,\sin(2\,z\,\kappa_{I}\,h_{I})\,,\\
&D_{S}^{[1]}=\,(h_{I}^{2}-\chi)\,\cos(2\,z\,\kappa_{I}\,h_{I})
-(h_{I}^{2}+\chi)\,\cosh(t\,h_{I}\\
&~~~~~~~~+2\,z\,\kappa_{R}\,h_{I})+\,(2\,b\,h_{I}+4\,m\,h_{I})\,\sin(2\,z\,\kappa_{I}\,h_{I})\\
&~~~~~~~~-4\,n\,h_{I}\,\sinh(t\,h_{I}+2\,z\,\kappa_{R}\,h_{I})\,,
\end{aligned}\]
 and for the periodic wave,
 \begin{equation}\label{GNLShi}
\begin{aligned}
& E_{P}^{[1]}=&d\,(1+8\,n\,\frac{G_{P}^{[1]}+i\,H_{P}^{[1]}}{D_{P}^{[1]}})\,e^{i\,\rho}
\end{aligned}
\end{equation}
with
\[
\begin{aligned}
&G_{P}^{[1]}=-2\,n\,\cos(t\,h_{R}+2\,z\,\kappa_{R}\,h_{R})+2\,n\,\cosh(2\,z\,\kappa_{I}\,h_{R})\\
&~~~~~~~~~+\,h_{R}\,\sin(t\,h_{R}+2\,z\,\kappa_{R}\,h_{R})\,,\\
&H_{P}^{[1]}=-\,(b+2\,m)\,\cos(t\,h_{R}+2\,z\,\kappa_{R}\,h_{R})+(b+2\,m)\\
&~~~~~~~~~\cosh(2\,z\,\kappa_{I}\,h_{R})+h_{R}\,\sinh(2\,z\,\kappa_{I}\,h_{R})\,,\\
&D_{P}^{[1]}=-\,(h_{R}^{2}-\chi)\,\cos(t\,h_{R}+2\,z\,\kappa_{R}\,h_{R})
-(h_{R}^{2}+\chi)\\
&~~~~~~~~~\cosh(2\,z\,\kappa_{I}\,h_{R})-4\,n\,h_{R}\,\sin(t\,h_{R}+2\,z\,\kappa_{R}\,h_{R})\\
&~~~~~~~~-(2\,b\,h_{R}+4\,m\,h_{R})\,\sinh(2\,z\,\kappa_{I}\,h_{R})\,.
\end{aligned}
\]

\textbf{Fig.~7} describes a soliton on the constant background, which propagates along $t$ direction.
This kind of soliton is called the antidark soliton firstly reported in the scalar NLS system
with the third-order dispersion~\cite{ANTI,ANTI_1}. Futher, as $d$ is approaching to zero, this soliton will turn
into a standard bright soliton. \textbf{Fig.~8} is plotted for the periodic waves propagating in $t$ direction with
the period $P=\frac{\pi}{h_{R}}$.

\section{RW-to-soliton conversion  }
\label{sec:2}\renewcommand{\theequation}{3.\arabic{equation}}\setcounter{equation}{0}
~~~~In this section, we investigate the transition of first-order localized wave
$|E_{1}(t, z)|^{2}$ from  the RW to the W-shaped traveling wave. Taking $\lambda\rightarrow-\frac{b}{2}+i\,d$ in Solution~(\ref{solution}), we
obtain the first-order RW  which is asymmetric with respect to $z$-axis in \textbf{Fig.~9}. Further, to convert the asymmetric RW into a soliton, the parameters $\omega$, $\tau$ and $b$ have to satisfy Eq.~(\ref{situation11}). Unlike the cases in Section~2, such transition is only controlled by three free parameters due to the fixed values of the real ($m=-\frac{b}{2}$) and imaginary ($n=d$) parts of the eigenvalue. In addition, Expression~(\ref{Rouge_kongshi}) has the forms of rational functions. Note that the rational solutions of the modified NLS equation have been found that they can describe various soliton states including a paired
bright-bright soliton, single soliton, a paired bright-grey soliton, a paired bright-black soliton, which is induced by self-steepening and self phase modulation through tuning the values of the corresponding  free parameters~\cite{HJS,HJS_1}. Hereby, we display that the rational solution of System~(\ref{GNLS_MB}) characterizes the W-shaped soliton which is plotted in \textbf{Fig.~10}. Different from the wave in Figs.~3 and 4, the soliton in Fig.~10 has only one (two), infinitely stretched, peak (valleys) without oscillatory tails. It is  worth noting that the W-shaped soliton can be also converted by the periodic wave in Fig.~8 when the period is infinite ($h_{R}\rightarrow0$). In other words, there are two
orders to obtain the W-shaped soliton: one is the breather$\rightarrow$RW$\rightarrow$W-shaped soliton,  and the other is the breather$\rightarrow$periodic wave$\rightarrow$W-shaped soliton.
\begin{equation}\label{Rouge_kongshi}
\begin{aligned}
&E_{W}^{[1]}=d\,\left(1- \frac{\Upsilon_{1}}{\Upsilon_{2}}\right)\,e^{\Upsilon_{3}}
\end{aligned}
\end{equation}
with
\[
\begin{aligned}
&\Upsilon_{1}=4 d (-8 \omega ^3 (4 b^3 d z+6 b d^3 z-3 (b-d) (b+d) (d t+1))\\
   &~~~~~~ +12 b \omega ^2 (4 b^3 d z+6 b d^3 z-3 (b-d) (b+d)(d t+1))\\
   &~~~~~~ -3 (b^5+3  b^3 d^2-4 b d^4) (d t+1)+2 b d z ((11 b^3+30)\\
   &~~~~~~~  d^2+2 b^2 (b^3-5)+12 b d^4)-2 \omega (2 d z ((17 b^3+6) d^2+\\
   &~~~~~~~ 6 b^2 (b^3-1)+12 b d^4)-3 (3 b^4+b^2 d^2-4 d^4) (d t+ 1)))\\
   &~~~~~~~ (4 b^6 z-3 b^5 (t+8 \omega  z)+2 b^4 (11 d^2 z+3 \omega  (3 t+8 \omega  z))\\
   &~~~~~~ -b^3 (d^2 (9 t+68 \omega  z)+4 \omega ^2 (9 t+8 \omega  z)+20 z)  +6 b^2 \\
   &~~~~~~~(4 d^4 z+d^2 \omega  (t+12 \omega  z)+4 \omega  (t \omega ^2+z))+12 b d^2 (d^2 \\
   &~~~~~~~(t-4 \omega  z)+3 t \omega  ^2-4 \omega ^3 z+5 z) -24 d^2 \omega  (t (d^2+\omega ^2)\\
   &~~~~~~+z))\,,\\
   &\Upsilon_{2}=32 d^2 z^2 b^{12}-24 d z b^{11}-48 d^2 t z b^{11}+18 d^2 t^2 b^{10}+352\\
   &~~~~~~~ d^4 z^2 b^{10}+18 d t b^{10}+9 b^{10}-320 d^2 z^2 b^9-204 d^3 z b^9-\\
   &~~~~~~~408 d^4 t z b^9 +54 d^2 b^8+108 d^4 t^2 b^8+1352 d^6 z^2 b^8+108\\
   &~~~~~~~ d^3 t b^8+120 d z b^8+240 d^2 t z b^8-800 d^4 z^2 b^7-444 d^5 z \\
   &~~~~~~~b^7-888 d^6 t z b^7+9 d^4 b^6+18  d^6 t^2 b^6+2112 d^8 z^2 b^6+\\
   \end{aligned}
\]
\[
\begin{aligned}
&~~~~~~~800 d^2 z^2 b^6+18 d^5 t b^6 +3360 d^6 z^2 b^5+96 d^7 z b^5+192  \\
   &~~~~~~~d^8 t  z b^5-216 d^6 b^4-432 d^8 t^2 b^4+1152 d^{10} z^2  b^4-4800 \\
   &~~~~~~~ d^4 z^2 b^4-432 d^7 t b^4 -1560 d^5 z b^4  -3120 d^6 t z b^4+5760 \\
   &~~~~~~~d^8 z^2 b^3+576 d^9 z b^3+1152 d^{10} t z b^3+144 d^8 b^2+288 d^{10} \\
   &~~~~~~~ t^2 b^2 +7200d^6 z^2 b^2+288 d^9 t b^2   +1440 d^7 z b^2+2880 d^8\\
   &~~~~~~~ t z b^2-192 (9 (b-d)^2 (2 d t (d t+1)+1) (b+d)^2+8 b^2 d^2 \\
   &~~~~~~~(2 b^2+3 d^2 )^2 z^2-12 d (2 b^5+d^2 b^3-3 d^4 b) (2 d t+1) z) \omega ^5 \\
   &~~~~~~~  b-32 (9 (b-d)^2 (5 b^2+8 d^2) (2 d t (d t+1)+1) (b+d)^2\\
   &~~~~~~-12(b-d) d (24 b d^4+(31 b^3+12) d^2+b^2 (10 b^3-7)) \\
   &~~~~~~~ (2 d t+1) z (b+d)+8 b d^2 (2 b^2+3 d^2)(24 b d^4+(31 b^3+\\
   &~~~~~~~ 24)d^2+2 b^2 (5 b^3-7)) z^2) \omega ^3 b-4 (9 (b-d)^2 (b^2+4 d^2)\\
   &~~~~~~~(3 b^2 +4 d^2) (2 d t (d t+1)+1) (b+d)^2-12 (b-d) d (48\\
   &~~~~~~~b d^6+8 (10 b^3+9) d^4+b^2 (41 b^3+16) d^2+6 b^4 (b^3-3)) \\
   &~~~~~~~(2 d t+1) z (b+d)+8 d^2 (12 b d^4+(11 b^3+30) d^2+2 b^2 \\
   &~~~~~~~(b^3-5)) (12 b d^4+(17  b^3+6) d^2+6 b^2 (b^3-1)) z^2) \omega  b+\\
   &~~~~~~~64 (9 (b-d)^2 (2 d t (d t+1)+1) (b+d)^2+8 b^2 d^2 (2 b^2+3\\
   &~~~~~~~ d^2)^2  z^2-12 d (2 b^5+d^2 b^3-3 d^4 b) (2 d t+1) z) \omega ^6+16 (9 \\
   &~~~~~~~(b-d)^2 (15 b^2+8 d^2) (2 d t (d t+1)+1) (b+d)^2-12 (b-\\
   &~~~~~~~d) d (24 b d^4+(61 b^3+6) d^2+6 b^2 (5 b^3-1)) (2 d t+1) z (b\\
   &~~~~~~+d)+8 b d^2 (2 b^2+3 d^2) (24 b d^4+(61 b^3+12) d^2+6 b^2 (5\\
   &~~~~~~~ b^3-2)) z^2) \omega ^4+4 (9 (b-d)^2 (15 b^4+48 d^2 b^2+16 d^4) (2 \\
   &~~~~~~~d t (d  t+1)+1) (b+d)^2-12 (b-d) d (48 b d^6+8 (22 b^3\\
   &~~~~~~~+3) d^4+3 b^2 (47 b^3+28) d^2+6 b^4 (5 b^3-8)) (2 d t+1) \\
   &~~~~~~~z (b+d)+8 d^2 (144 b^2 d^8+48 b (13 b^3+3) d^6+(775 b^6+\\
   &~~~~~~~600 b^3+36) d^4+12 b^2 (31 b^6+4 b^3-6) d^2+12 b^4 (5 b^6\\
   &~~~~~~-16 b^3+3)) z^2) \omega ^2\,,\\
   &\Upsilon_{3}=i (z (-b^2+((b^4-12 b^2 d^2+6 d^4) ((b-2 \omega ) ((b-2 \omega )^2+\\
   &~~~~~~~4 d^2)+4))/(6 (b-d) (b+d) (b-2 \omega ) ((b-2 \omega )^2+4 d^2))\\
   &~~~~~~~-4/(b-2 \omega )+2 d^2)+b t)\,.
\end{aligned}
\]
\section{Conclusions}
\label{sec:3}\renewcommand{\theequation}{4.\arabic{equation}}\setcounter{equation}{0}
~~~In this article,  we have shown that the first-order breather solution of System~(\ref{GNLS_MB}) can be converted into several types of
novel solitons including the W-shaped soliton  (multi-peak), M-shaped soliton, periodic wave and  antidark soliton. The transition condition~(\ref{situation11}), depending on the parameters $m$, $n$, $\omega$, $\tau$ and $b$, has been analytically presented. Additionally, we have exhibited that the first-order RW solution of System~(\ref{GNLS_MB}) can be converted into the W-shaped soliton with single peak and two valleys when the $\omega$, $\tau$ and $b$ satisfy the condition~(\ref{situation11}). It will be interesting to study the interactions among different kinds of nonlinear waves. We will give the details in another paper.

\vspace{7mm} \noindent {\bf Acknowledgements}
\vspace{3mm}

We express our sincere thanks to all the members of our discussion group for their valuable comments.
This work has been supported by the National Natural Science Foundation of China under Grant No.~11305060,   the Fundamental Research Funds of the Central Universities (No.~2015ZD16), by the Innovative Talents Scheme of North China Electric Power University, and by the higher-level item cultivation project of Beijing Wuzi University (No.~GJB20141001).

\begin{figure*}
\centering
\includegraphics[width=140 bp]{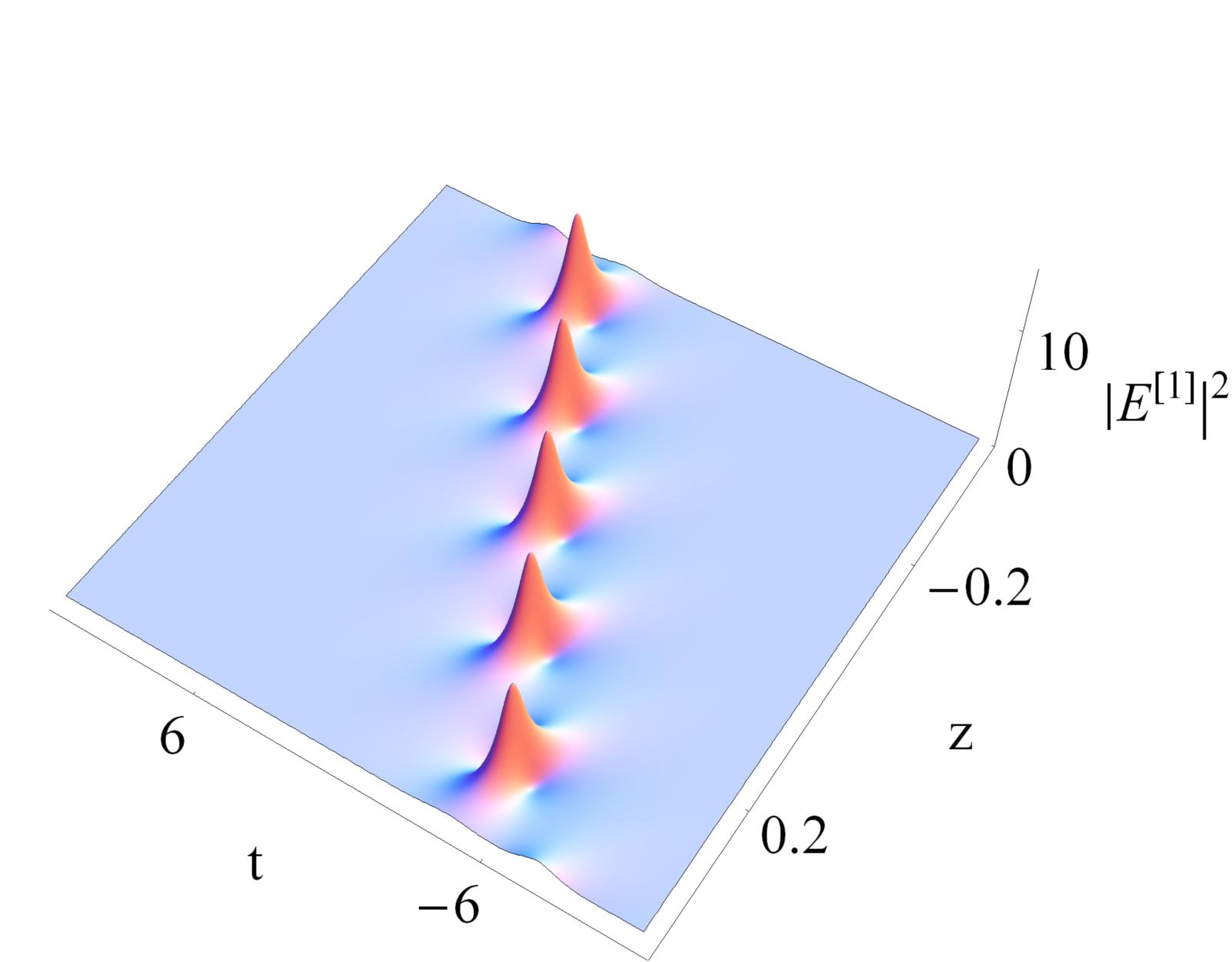}
\quad\quad
\includegraphics[width=140 bp]{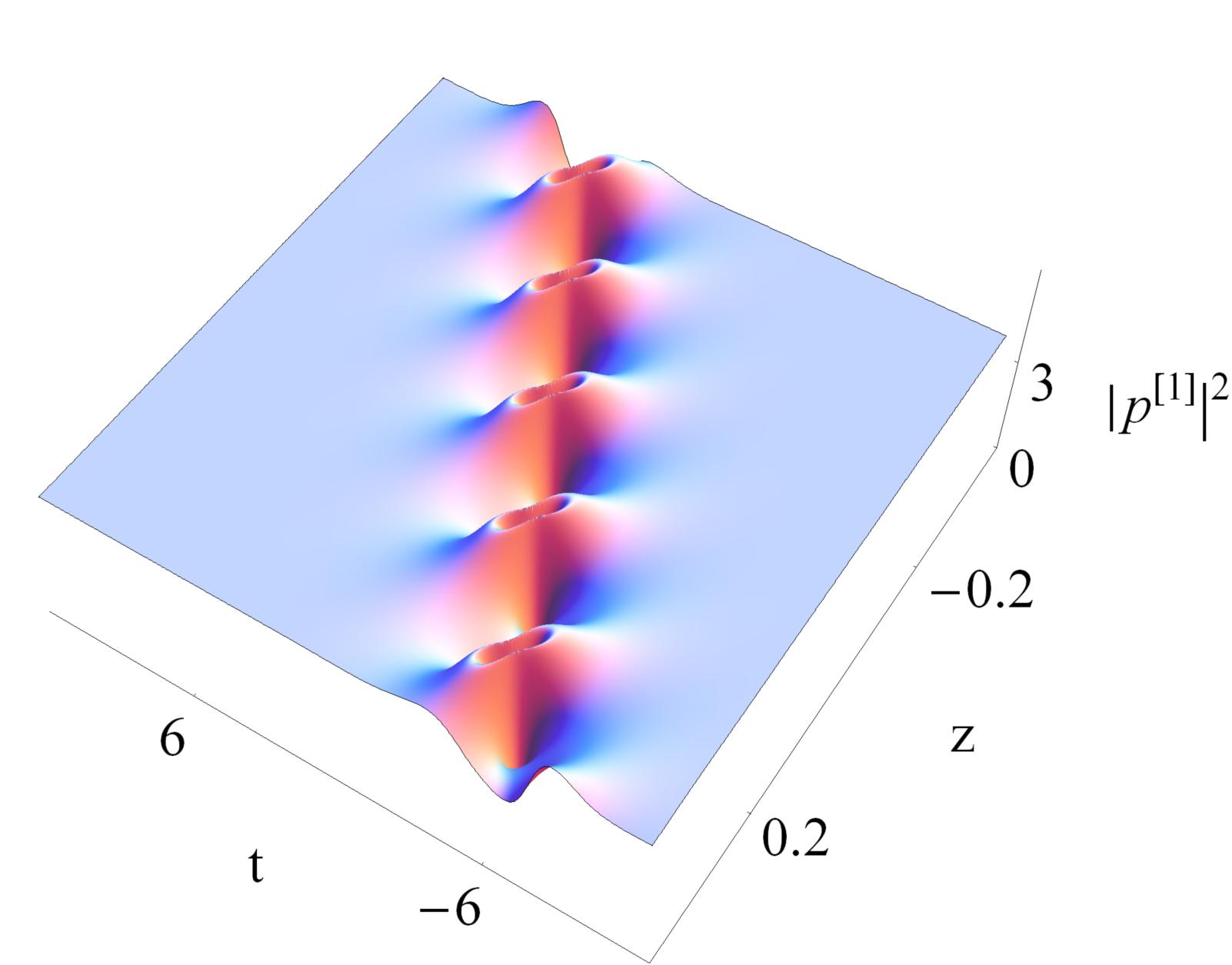}
\quad\quad
\includegraphics[width=140 bp]{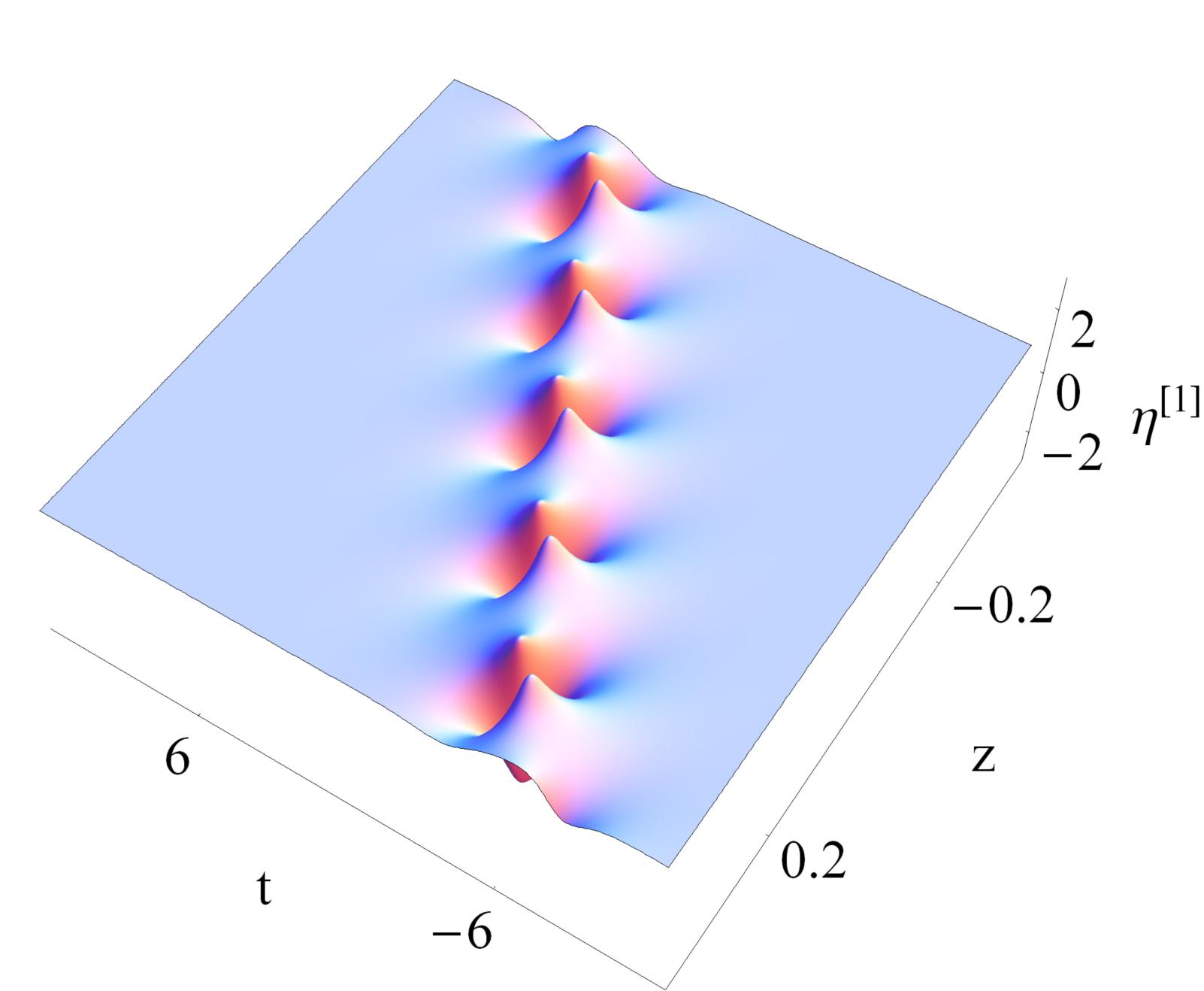}
{\footnotesize  Fig.~1(a) \hspace{4.2cm} Fig.~1(b) \hspace{4.8cm}Fig.~1(c)}
\caption{First-order breather solution of the GNLS-MB system ($E,p,\eta$) with $d=1,b=2,k_{1}=-k_{2}=1,\omega=\frac{1}{2},\tau=\frac{1}{2}$ and $\lambda_{1}=\lambda_{2}^{\ast}=-1.2+1\,i$}\label{figs:1}
\end{figure*}

\begin{figure*}
\centering
\includegraphics[width=130 bp]{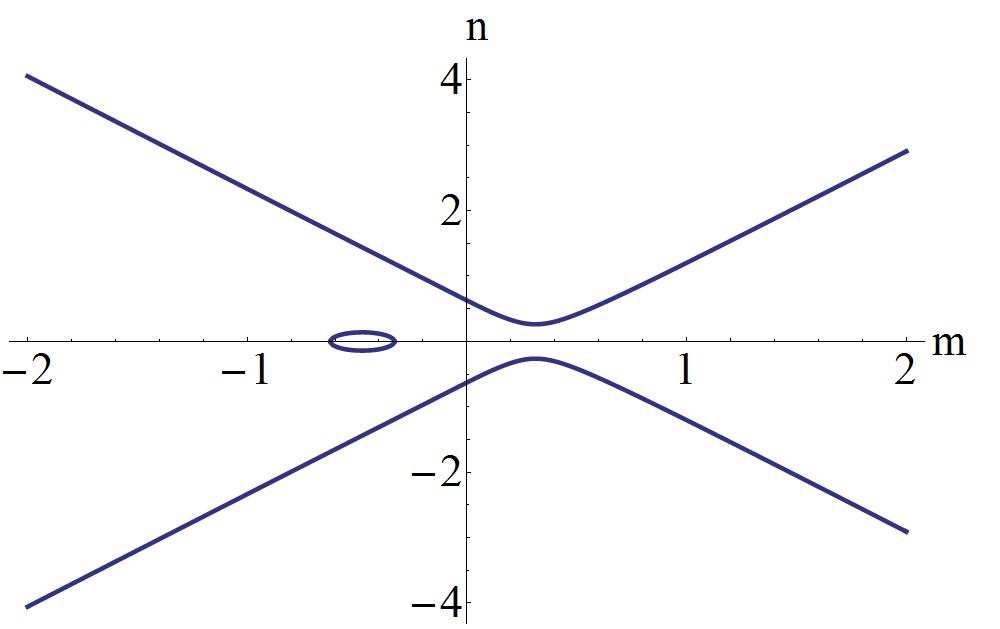}
\quad\quad
\includegraphics[width=130 bp]{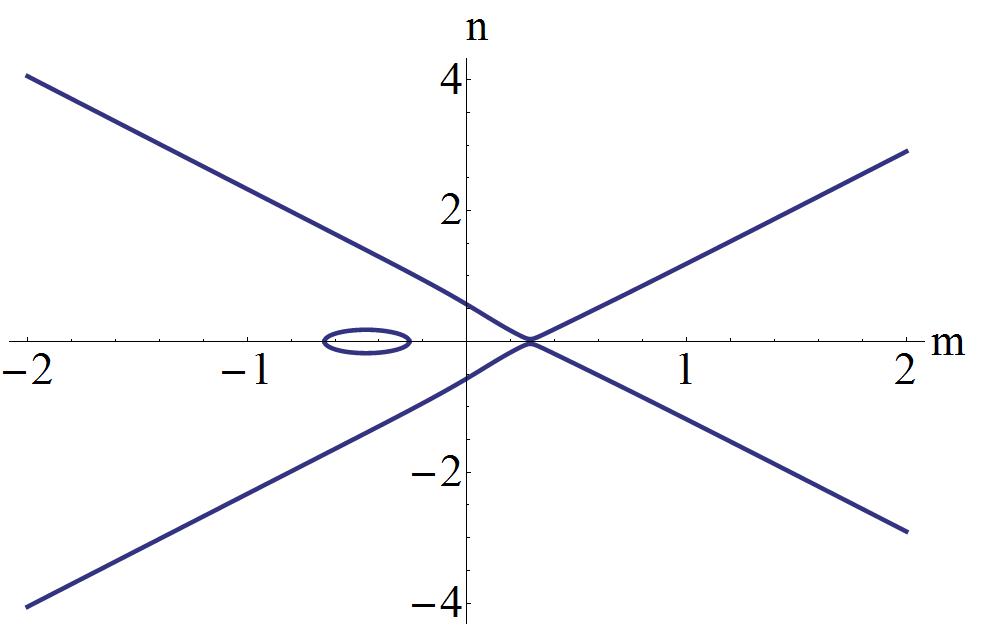}
\quad\quad
\includegraphics[width=130 bp]{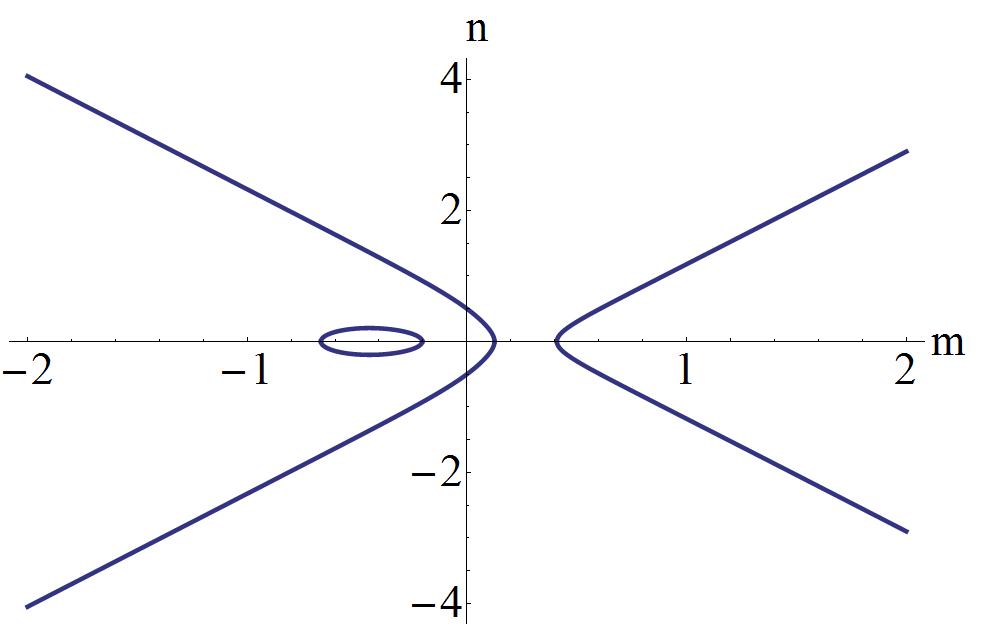}
{\footnotesize  \hspace{4cm} Fig.~2(a) \hspace{4cm} Fig.~2(b) \hspace{4cm}Fig.~2(c)}
\\\vspace{4mm}
\includegraphics[width=130 bp]{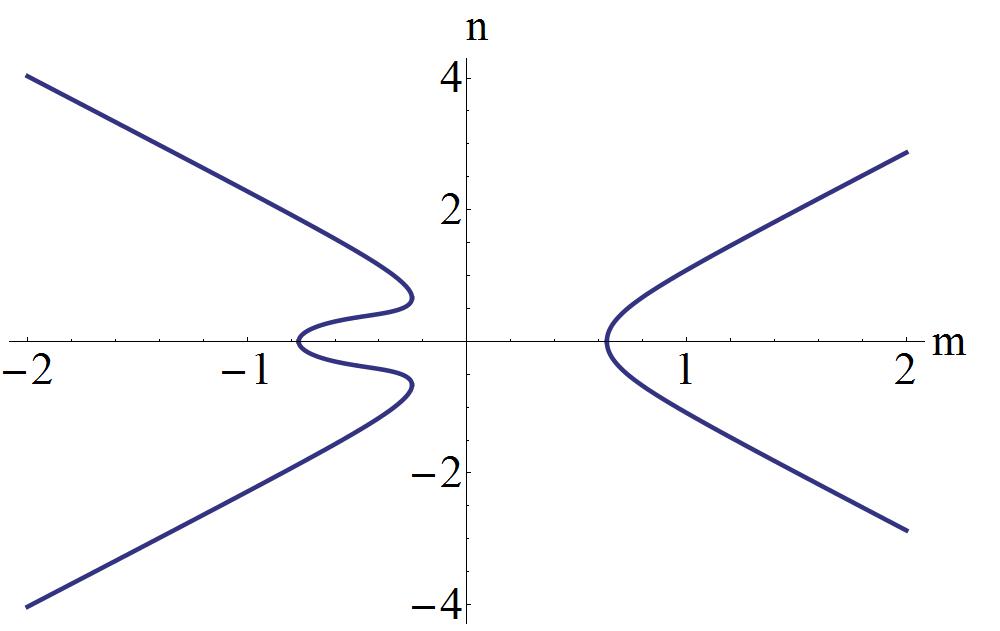}
\quad\quad
\includegraphics[width=130 bp]{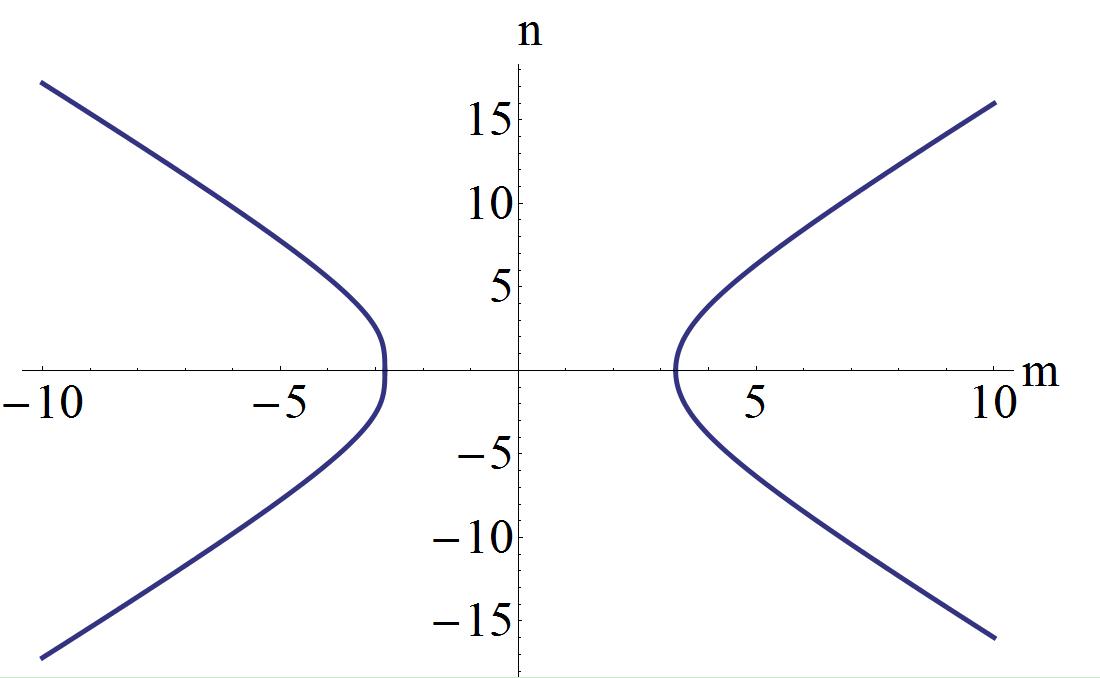}
\quad\quad
\includegraphics[width=130 bp]{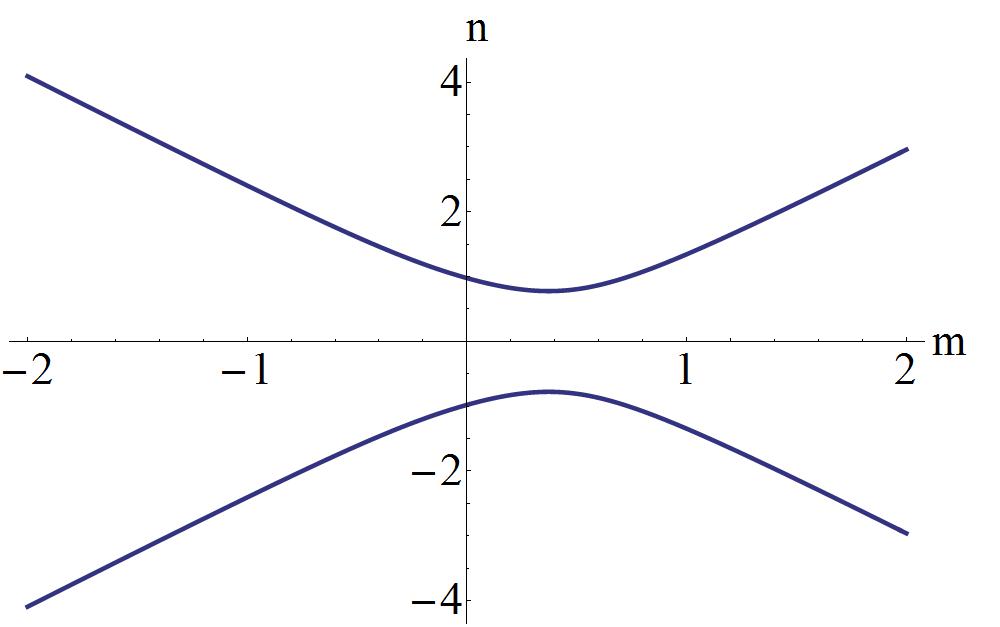}
{\footnotesize  \hspace{4cm} Fig.~2(d) \hspace{4cm} Fig.~2(e) \hspace{4cm}Fig.~2(f)}
\caption{\footnotesize Solutions of Eq.~(\ref{situation11}) on a complex plane of $\lambda_{1}=m+n\,i$.$\,d=1,b=2,\omega=\frac{1}{2}$. (a): $\tau=6$, (b): $\tau=3.8$, (c): $\tau=3$, (d):$\tau=1$, (e): $\tau=0.01$, (f): $\tau=-1$.}
\end{figure*}

\begin{figure*}
\centering
\includegraphics[width=150 bp]{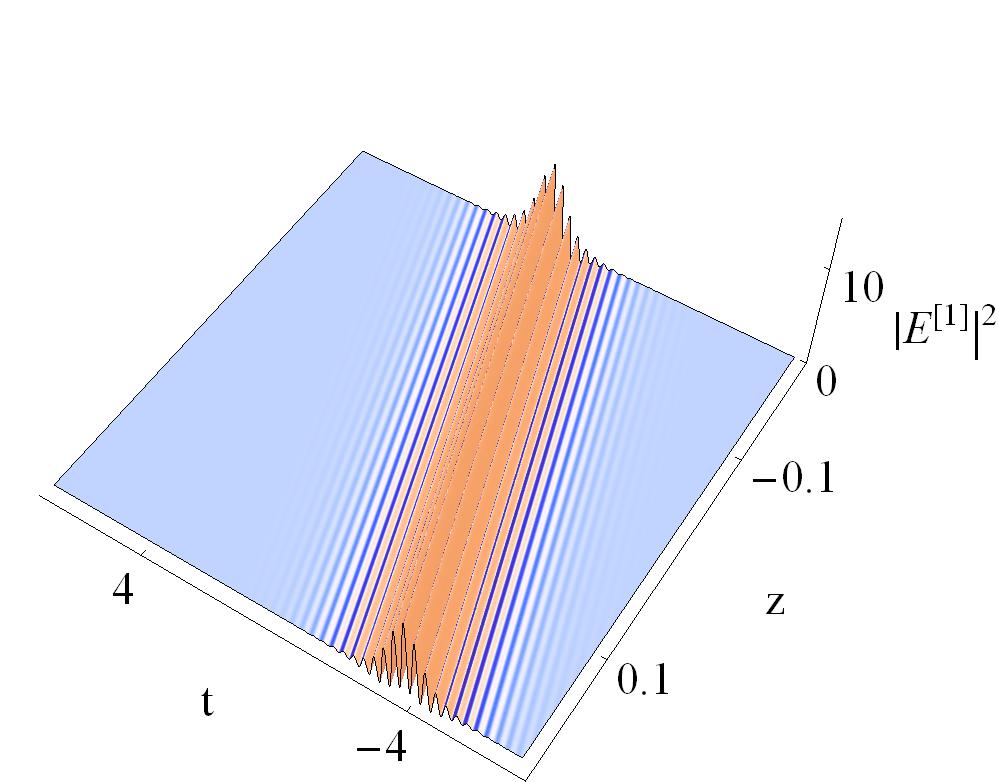}
\quad\quad
\includegraphics[width=100 bp]{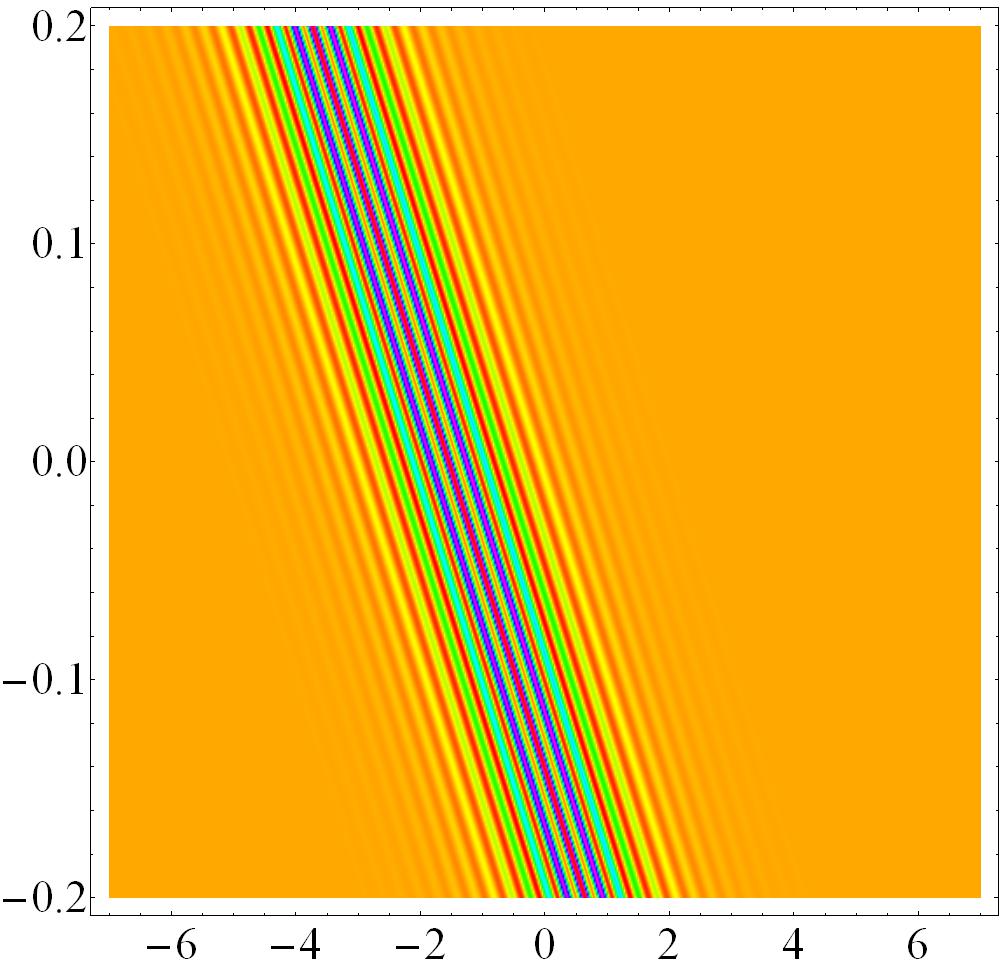}
\includegraphics[height=100 bp]{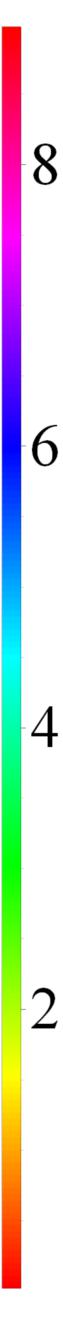}
\quad\quad\quad\quad
\includegraphics[width=150 bp]{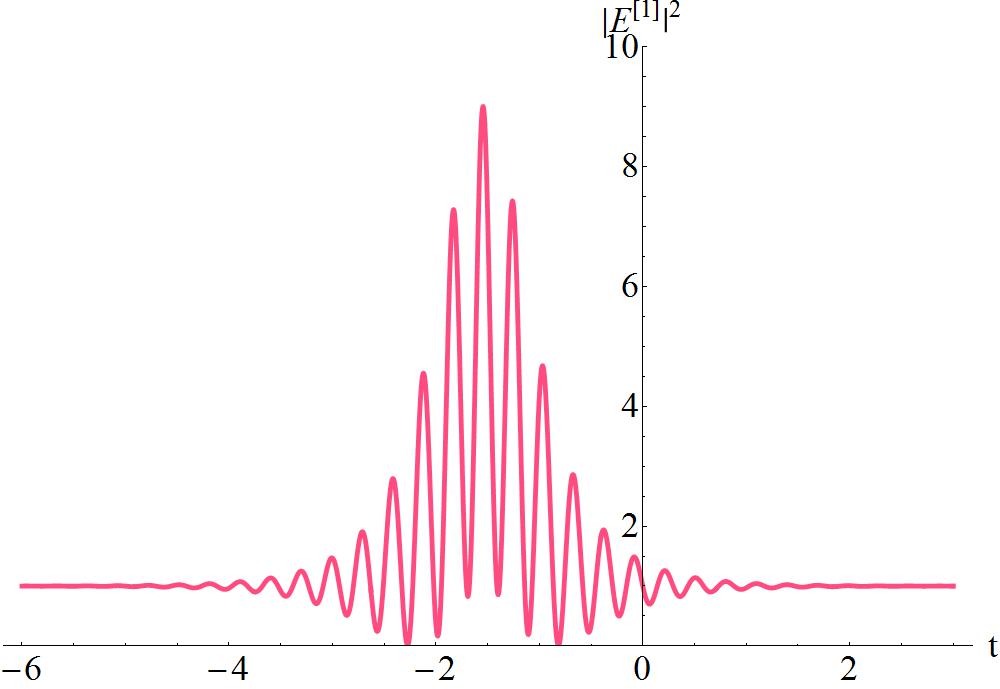}
{\footnotesize  \hspace{4.5cm} Fig.~3(a) \hspace{4cm} Fig.~3(b) \hspace{4.5cm}Fig.~3(c)}
\caption{\footnotesize A breather transformed into a multi-peak soliton with $d=1,b=2,k_{1}=-k_{2}=1,\omega=\frac{1}{2}$ and $\lambda_{1}=\lambda_{2}^{\ast}=9.6+i$. (b) is the contour plot of (a). (c) is the cross-sectional view of (a) at $z=0$. }
\end{figure*}

\begin{figure*}
\centering
\includegraphics[width=150 bp]{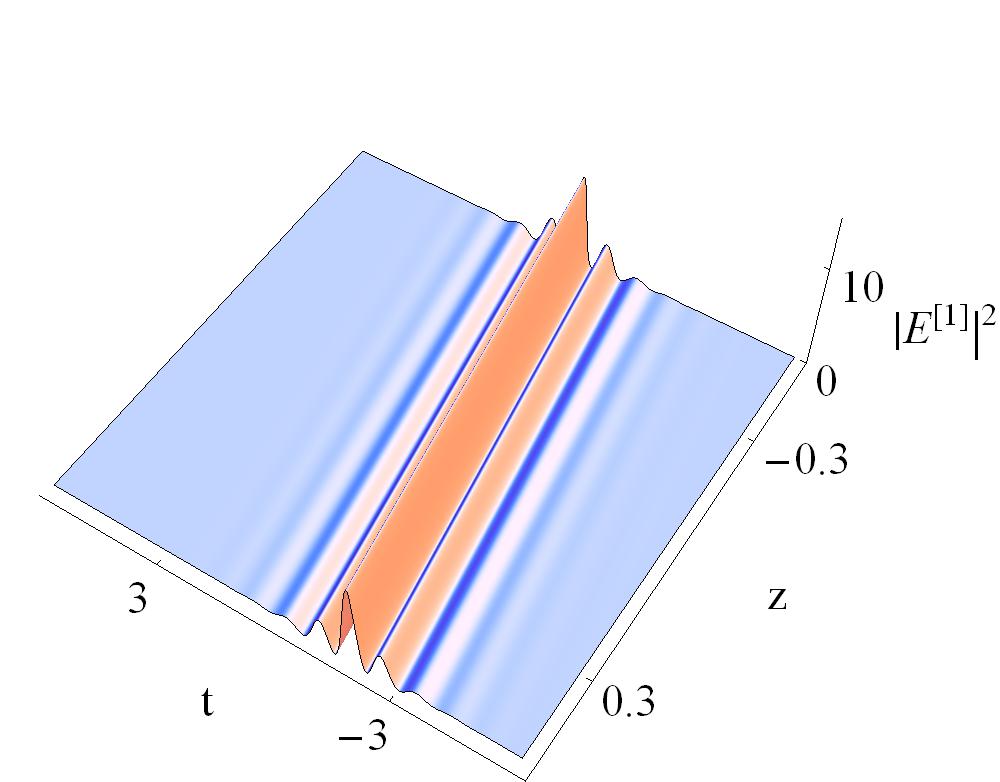}
\quad\quad
\includegraphics[width=100 bp]{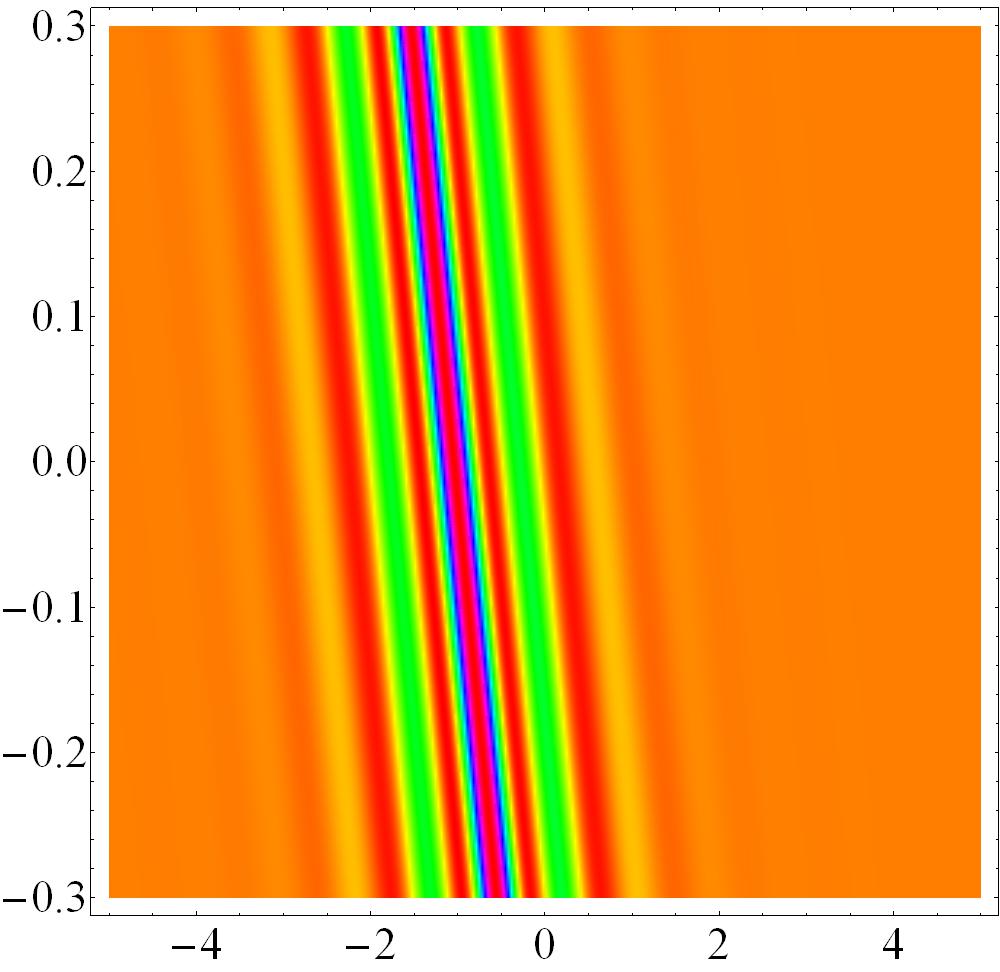}
\includegraphics[height=100 bp]{1-order-B-SW-21DF.jpg}
\quad\quad\quad\quad
\includegraphics[width=150 bp]{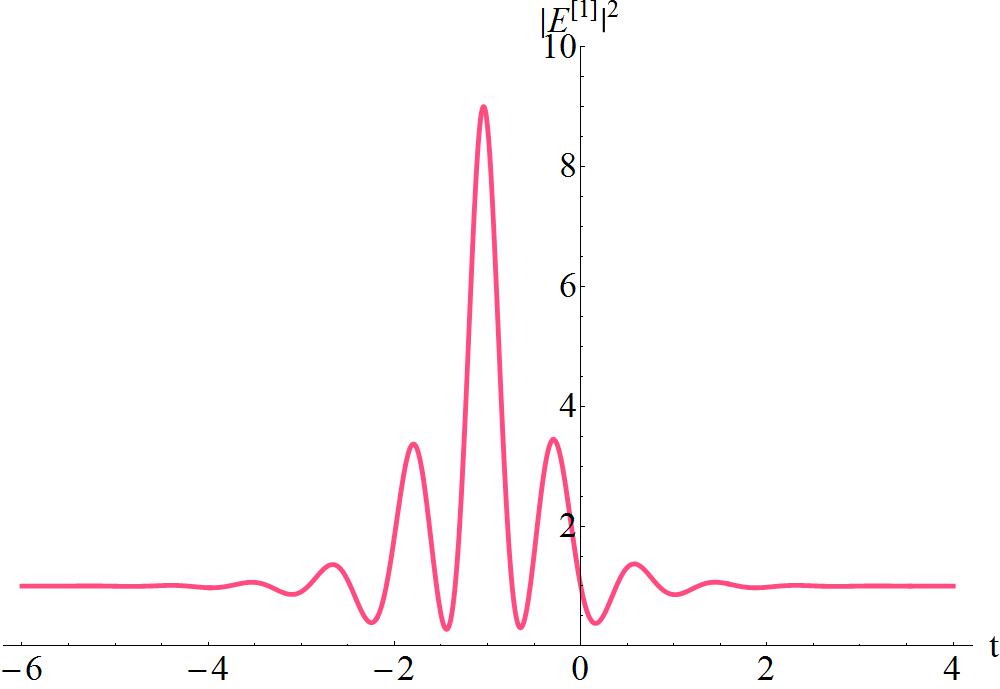}
{\footnotesize  \hspace{4.5cm} Fig.~4(a) \hspace{4cm} Fig.~4(b) \hspace{4.5cm}Fig.~4(c)}
\caption{\footnotesize A breather transformed into a multi-peak soliton with $d=1,b=2,k_{1}=-k_{2}=1,\omega=\frac{1}{2}$ and $\lambda_{1}=\lambda_{2}^{\ast}=2.5+i$. (b)is the contour plot of (a). (c) is the cross-sectional view of (a) at $z=0$. }
\end{figure*}

\begin{figure*}
\centering
\includegraphics[width=150 bp]{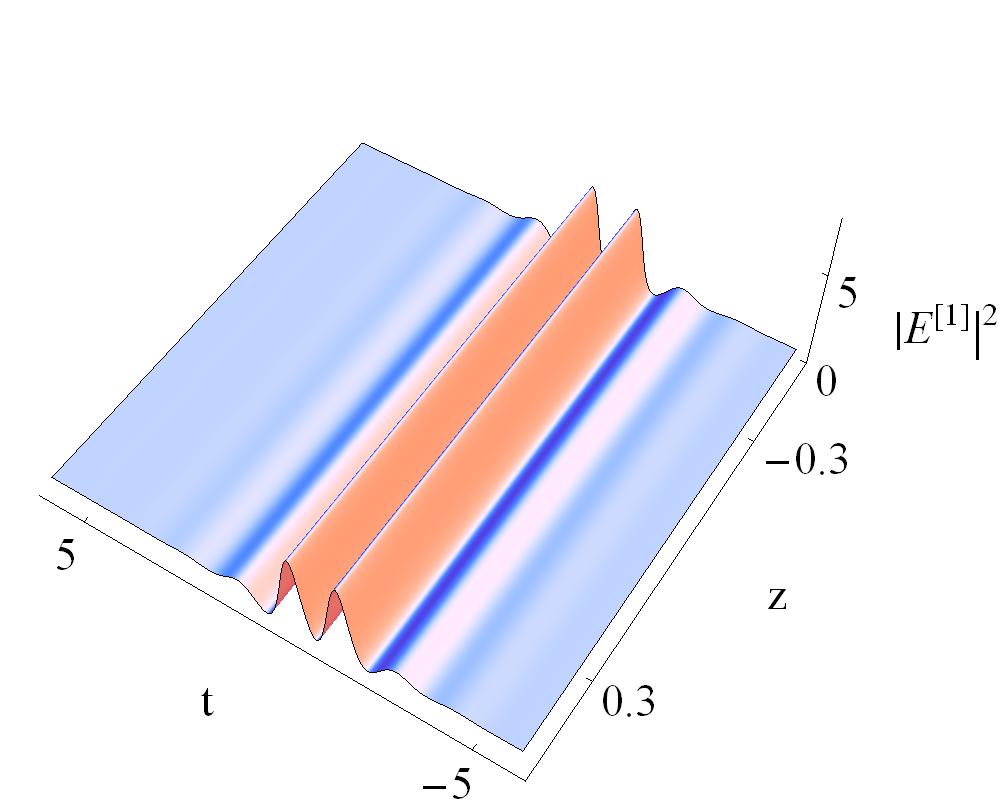}
\quad\quad
\includegraphics[width=100 bp]{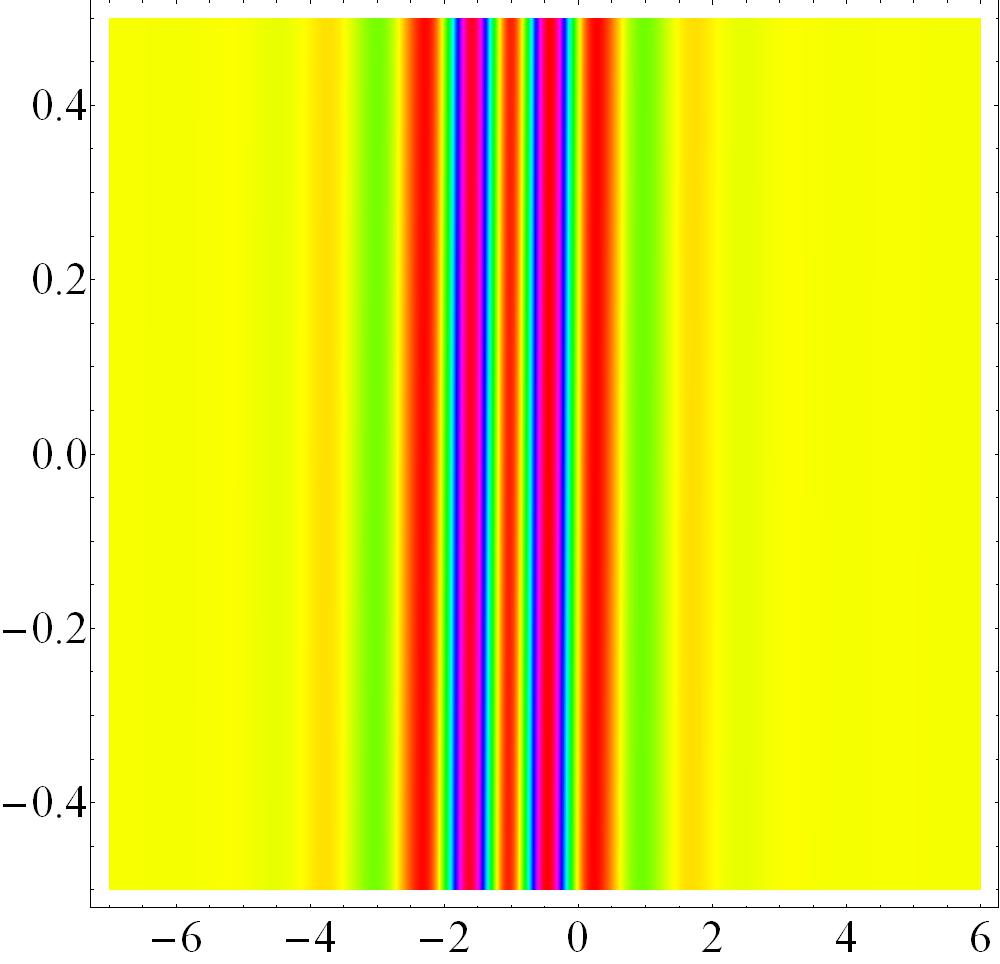}
\includegraphics[height=100 bp]{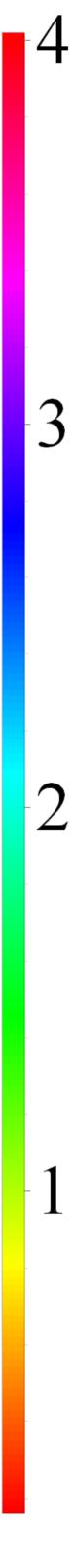}
\quad\quad\quad\quad
\includegraphics[width=150 bp]{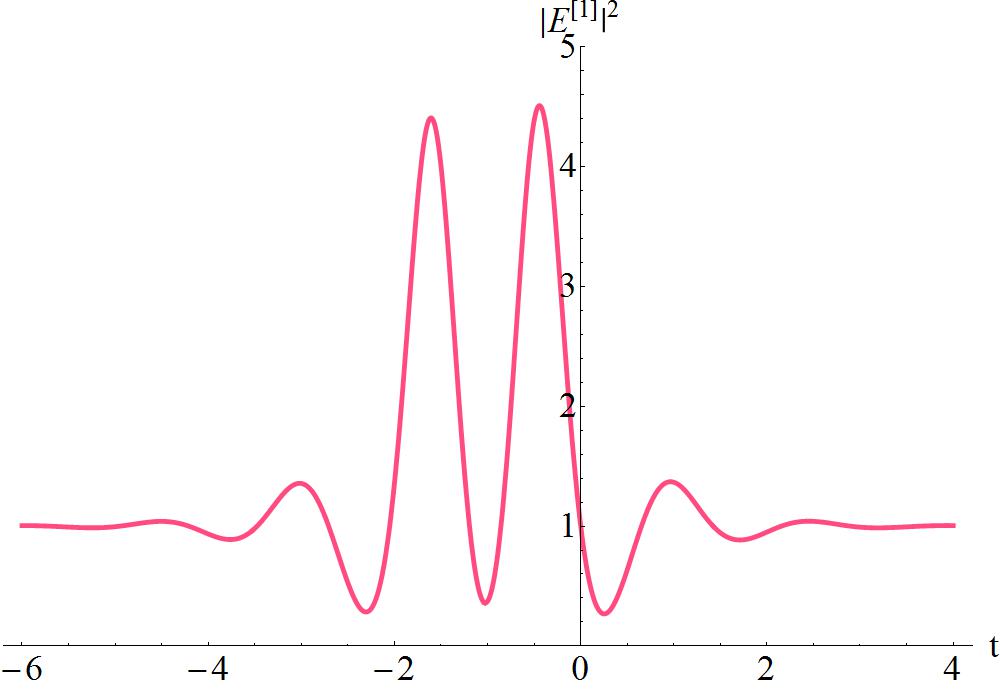}
{\footnotesize  \hspace{4.5cm} Fig.~5(a) \hspace{4cm} Fig.~5(b) \hspace{4.5cm}Fig.~5(c)}
\caption{\footnotesize A breather transformed into a M-shape soliton with $d=1,b=2,k_{1}=-k_{2}=1,\omega=\frac{1}{2}$ and $\lambda_{1}=\lambda_{2}^{\ast}=0.9+0.8\,i$. (b) is the contour plot of (a). (c) is the cross-sectional view of (a) at $z=0$.}
\end{figure*}

\begin{figure*}
\centering
\includegraphics[width=150 bp]{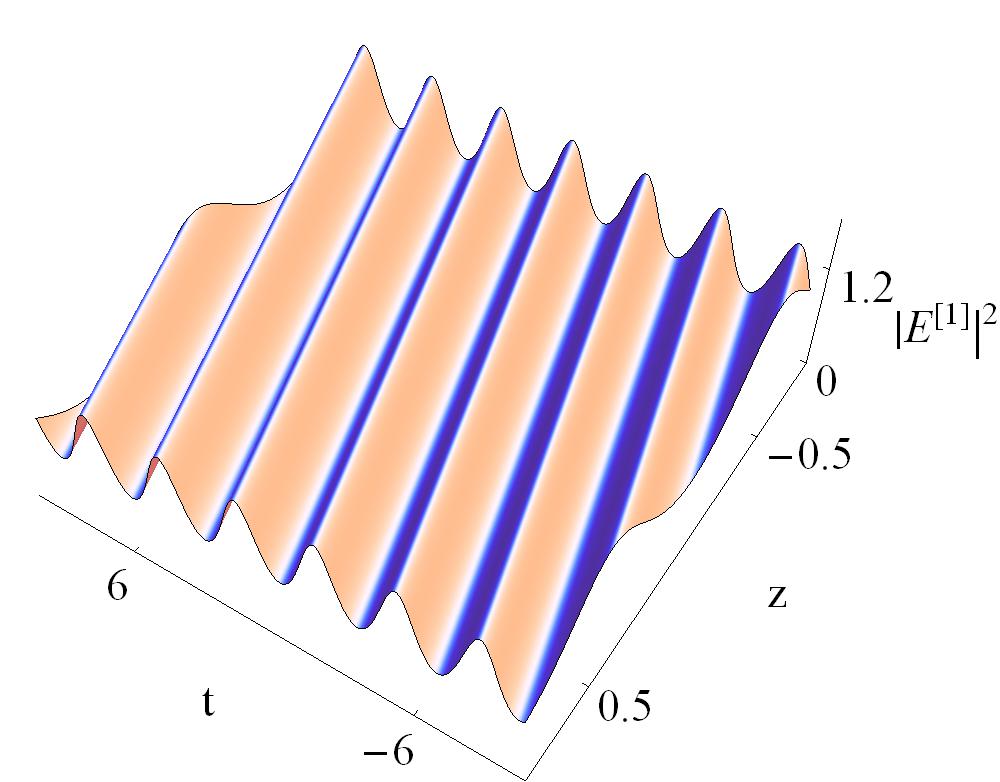}
\quad\quad
\includegraphics[width=100 bp]{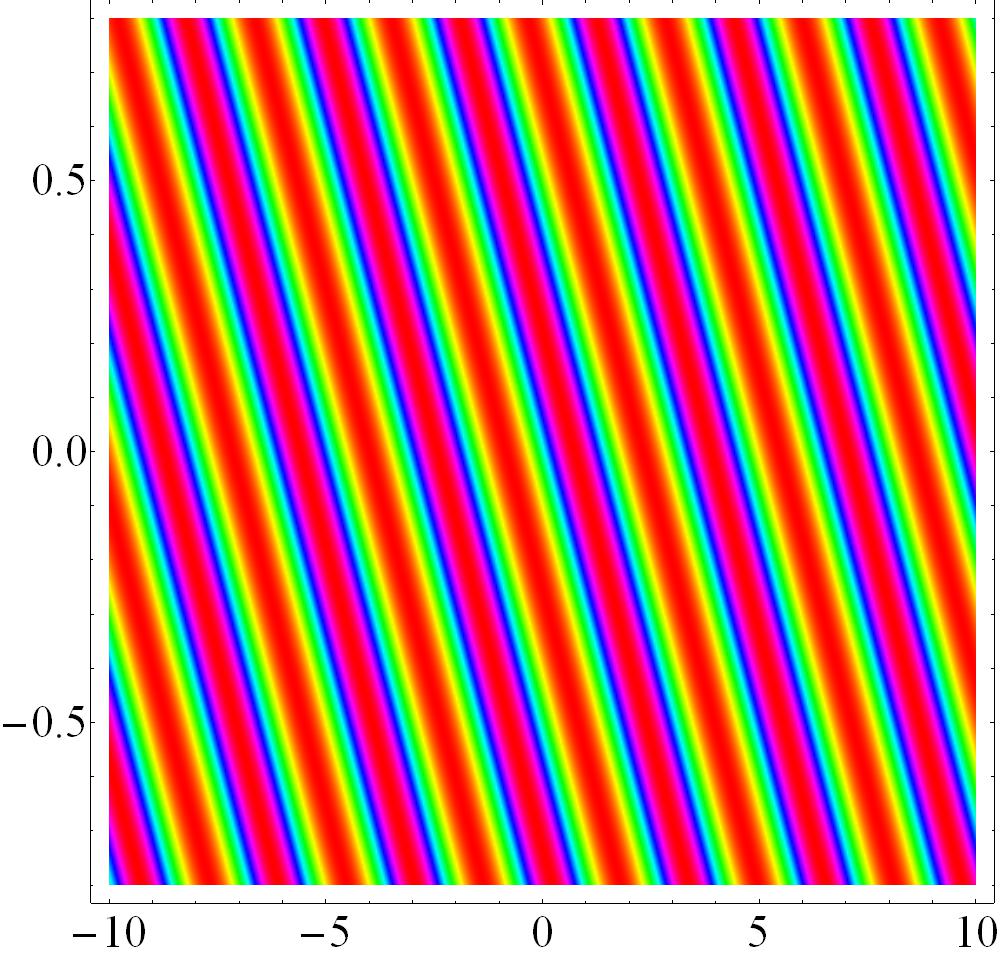}
\includegraphics[height=100 bp]{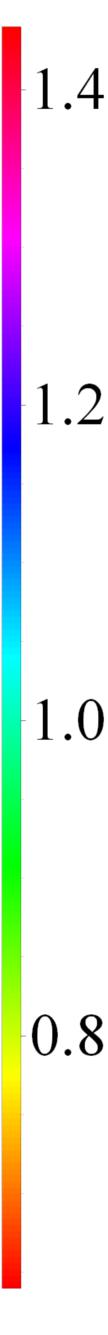}
\quad\quad\quad\quad
\includegraphics[width=150 bp]{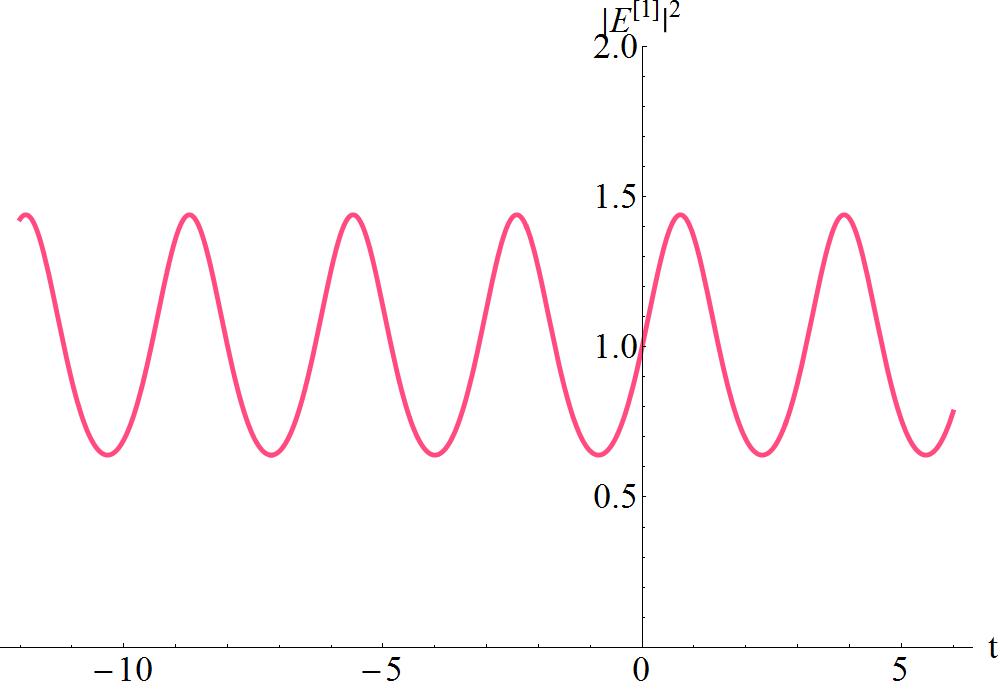}
{\footnotesize  \hspace{4.5cm} Fig.~6(a) \hspace{4cm} Fig.~6(b) \hspace{4.5cm}Fig.~6(c)}
\caption{\footnotesize A breather transformed into a multi-peak soliton (equal-amplitude) with $d=1,b=2,k_{1}=-k_{2}=1,\omega=\frac{1}{2}$ and $\lambda_{1}=\lambda_{2}^{\ast}=-1-0.1\,i$. (b) is the contour plot of (a).\,(c) is the cross-sectional view of (a) at $z=0$. }
\end{figure*}

\begin{figure*}
\centering
\includegraphics[width=150 bp]{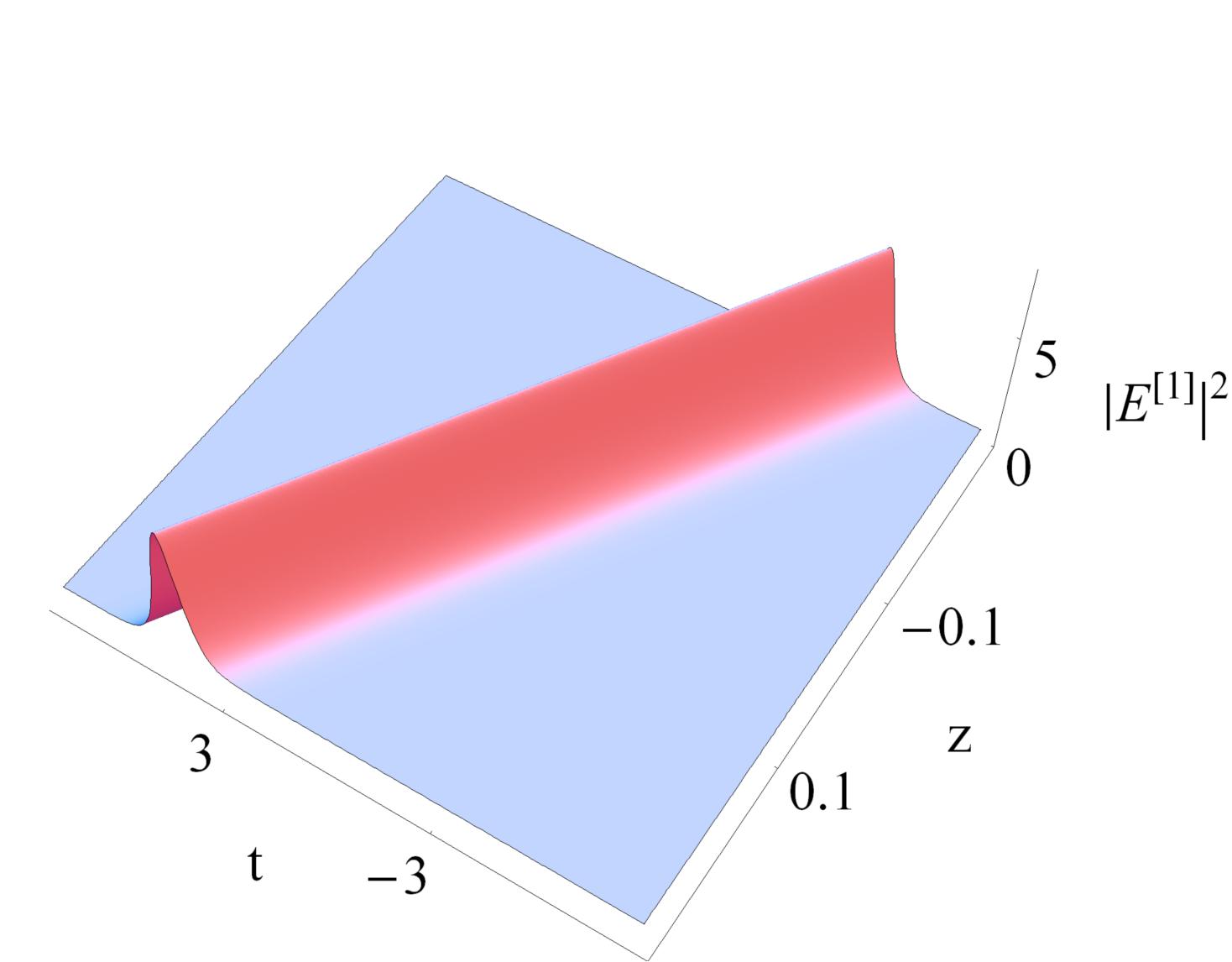}
\quad\quad
\includegraphics[width=100 bp]{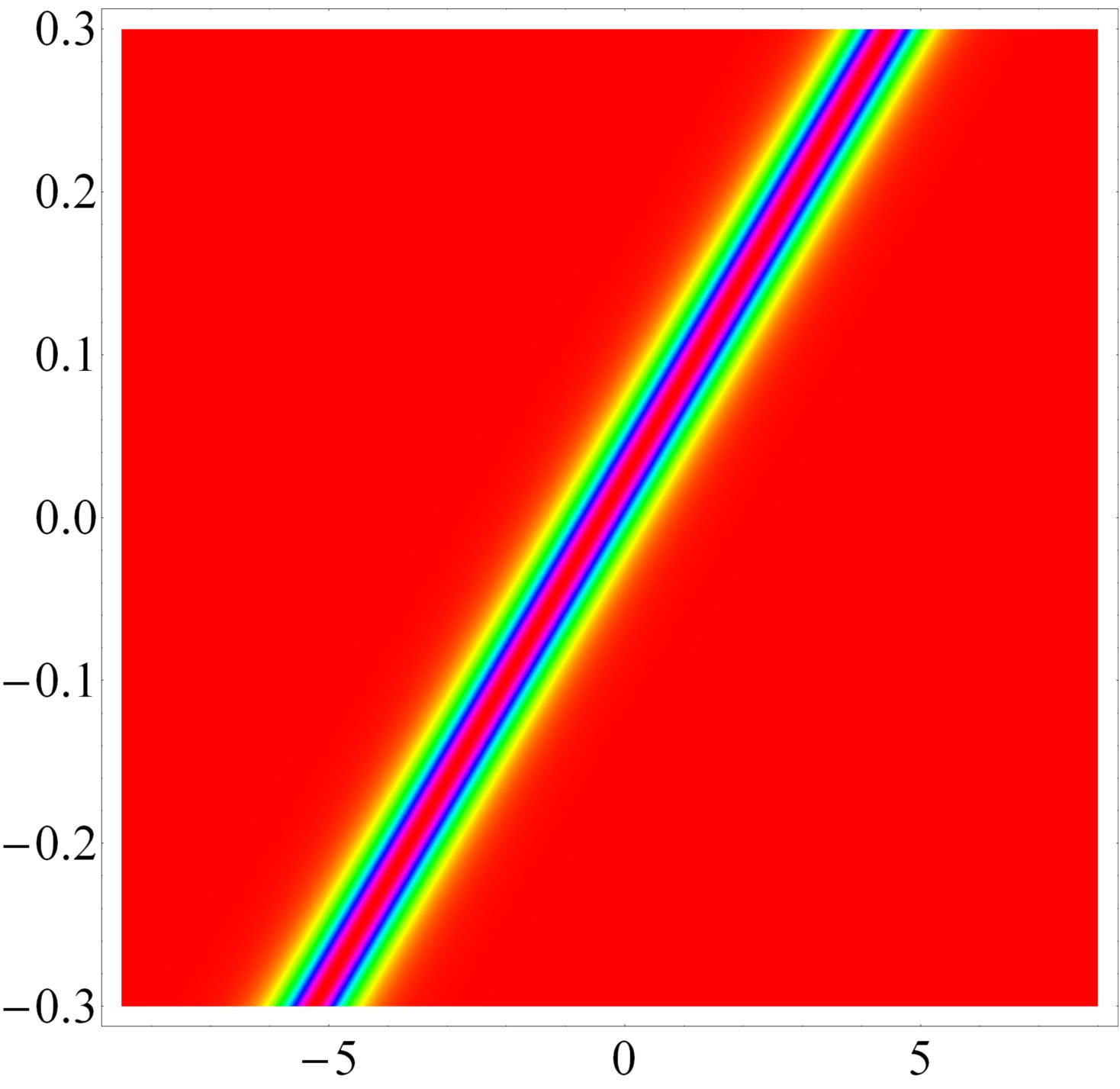}
\includegraphics[height=100 bp]{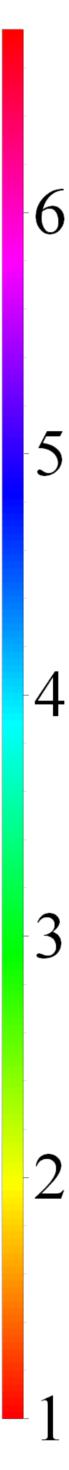}
\quad\quad\quad\quad
\includegraphics[width=150 bp]{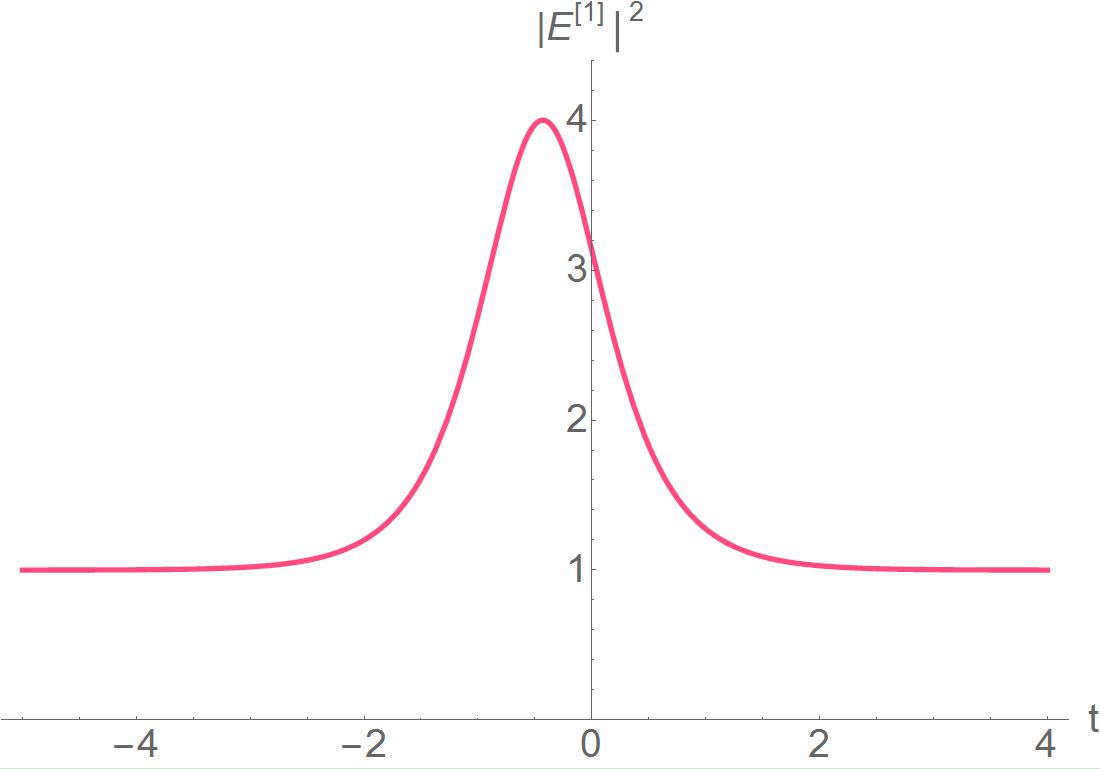}
{\footnotesize  \hspace{4.5cm} Fig.~7(a) \hspace{4cm} Fig.~7(b) \hspace{4.5cm}Fig.~7(c)}
\caption{\footnotesize A breather transformed into an  antidark soliton with $d=1,b=2,k_{1}=k_{2}=1,\omega=\frac{1}{2}$ and $\lambda_{1}=\lambda_{2}^{\ast}=-\frac{1}{2}\,b+1.5\,i$. (b) is the contour plot of (a). (c) is the cross-sectional view of (a) at $z=0$. }
\end{figure*}

\begin{figure*}
\centering
\includegraphics[width=150 bp]{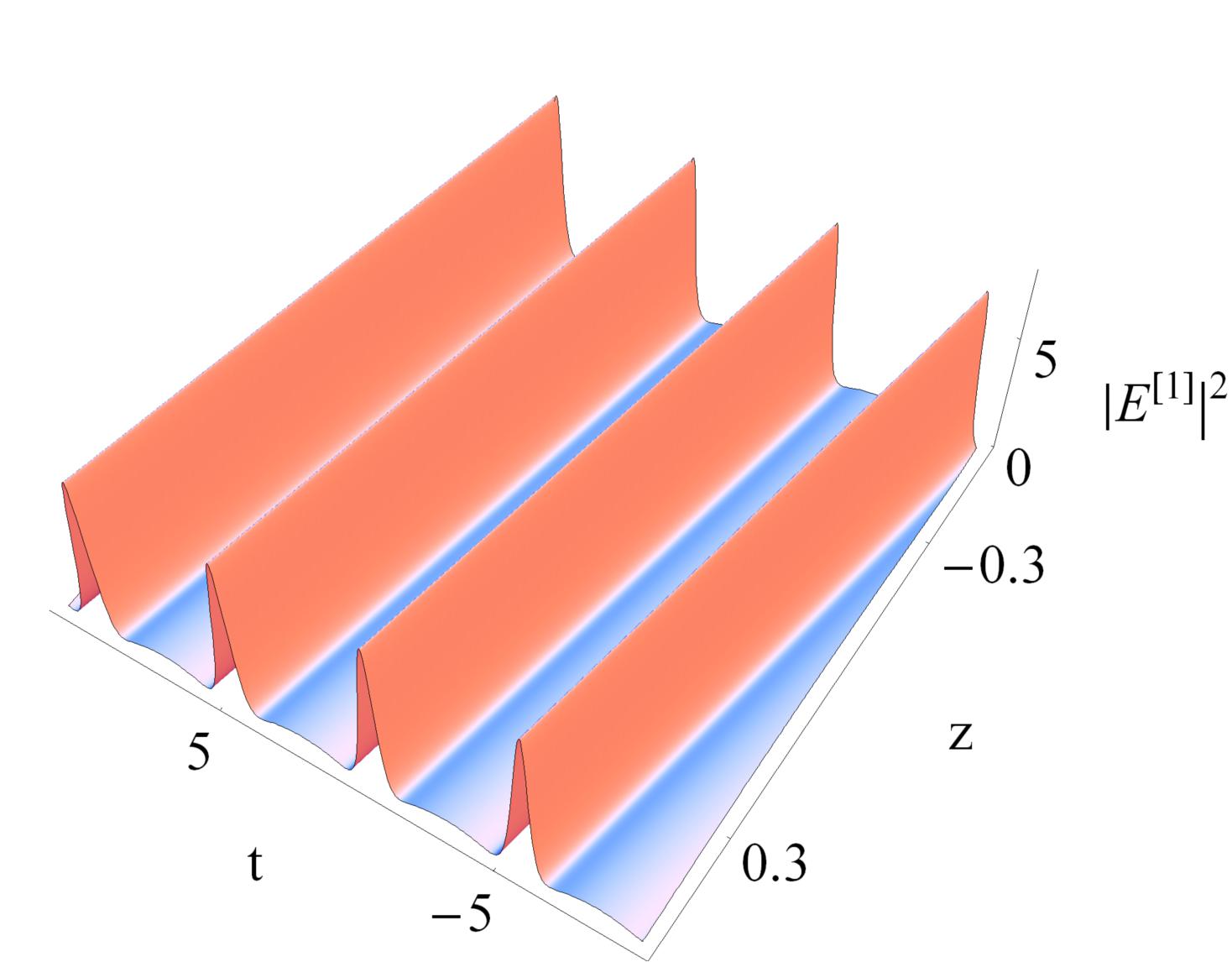}
\quad\quad
\includegraphics[width=100 bp]{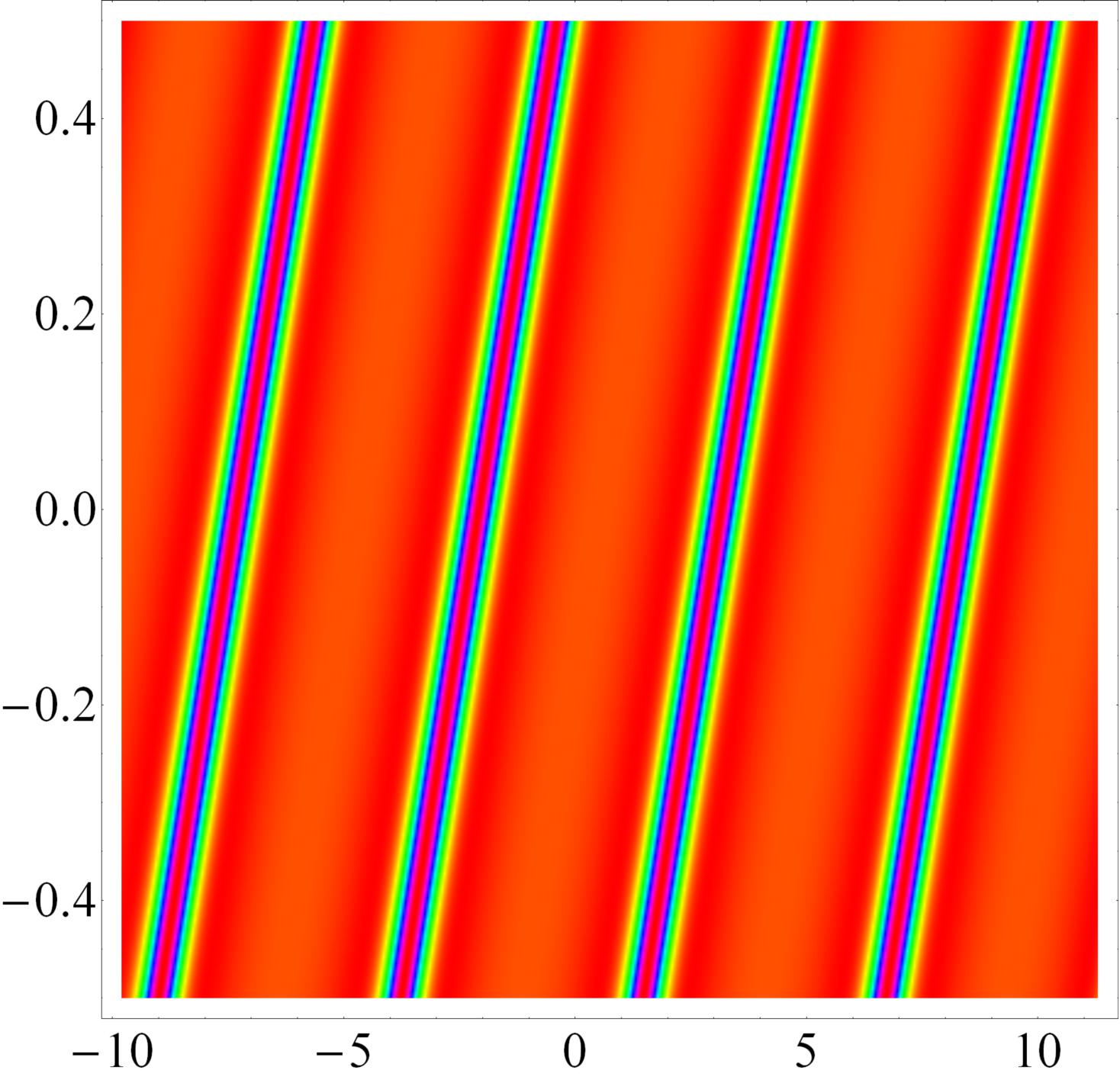}
\includegraphics[height=100 bp]{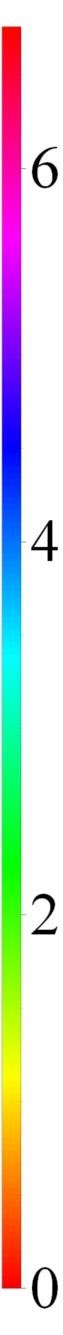}
\quad\quad\quad\quad
\includegraphics[width=150 bp]{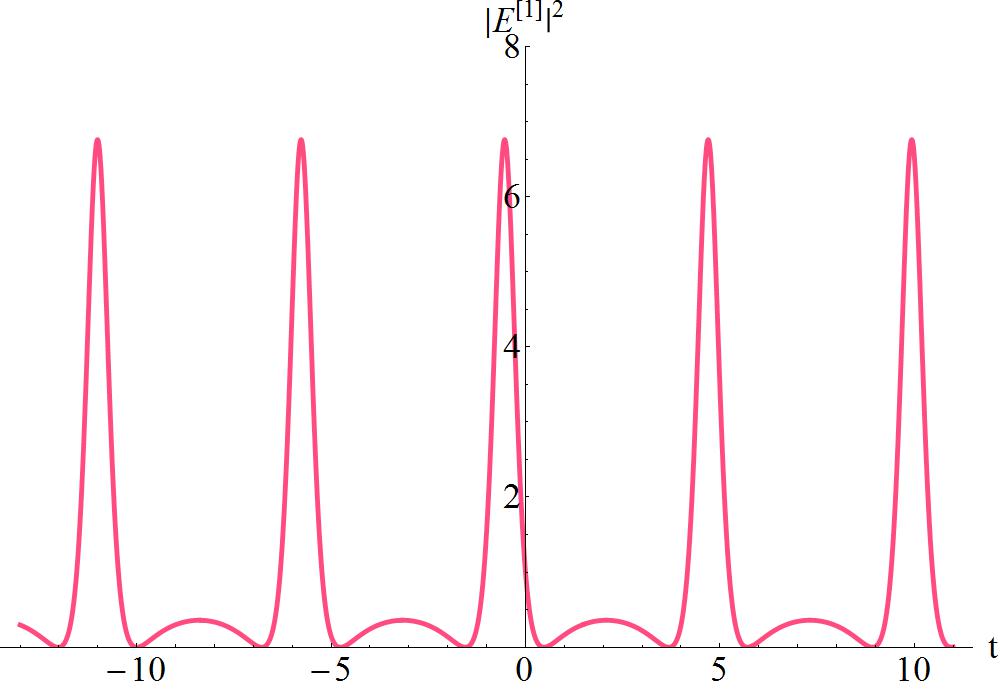}
{\footnotesize  \hspace{4.5cm} Fig.~8(a) \hspace{4cm} Fig.~8(b) \hspace{4.5cm}Fig.~8(c)}
\caption{\footnotesize A breather transformed into period wave with $d=1,b=2,k_{1}=-k_{2}=1,\omega=\frac{1}{2}$ and $\lambda_{1}=\lambda_{2}^{\ast}=-\frac{1}{2}\,b+0.8\,i$. (b) is the contour plot of (a). (c) is the cross-sectional view of (a) at $z=0$. }
\end{figure*}

\begin{figure*}
\centering
\includegraphics[width=150 bp]{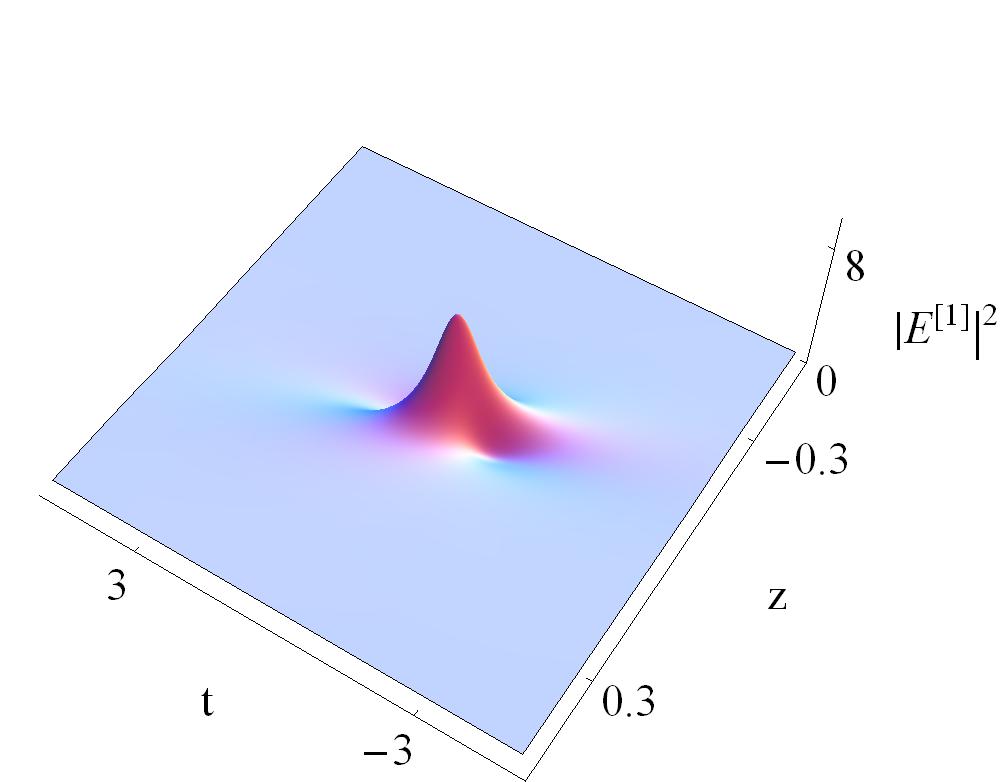}
\quad\quad\quad\quad\quad\quad\quad
\includegraphics[width=100 bp]{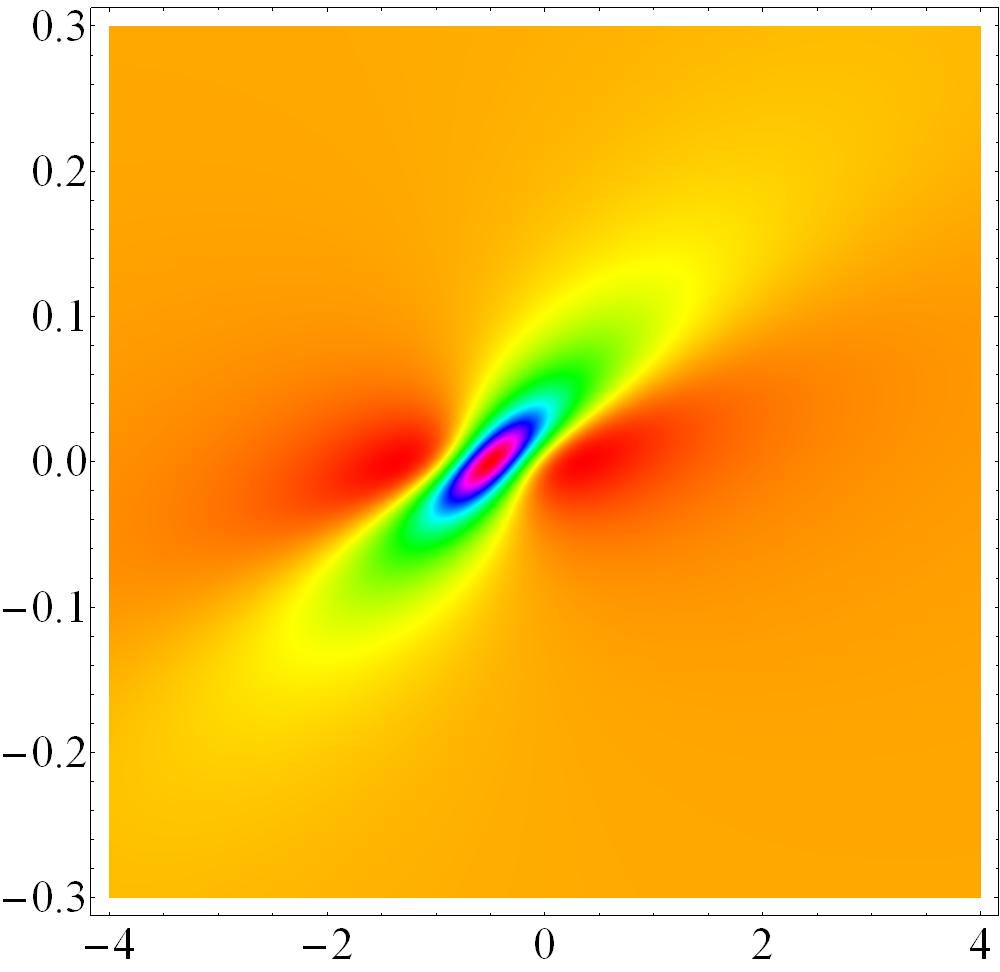}
\includegraphics[height=100 bp]{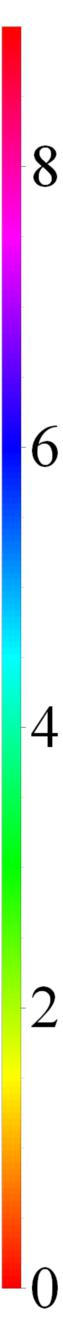}\\
{\footnotesize  Fig.~9(a) \hspace{6cm} Fig.~9(b) }
\caption{\footnotesize First-order  asymmetric RW solution with $d=1, b=2, k_{1}=-k_{2}=1,\omega=\frac{1}{2},\tau=\frac{1}{2}$ and $\lambda_{1}=\lambda_{2}^{\ast}=-\frac{1}{2}\,b+i\,d $.}
\end{figure*}

\begin{figure*}
\centering
\includegraphics[width=150 bp]{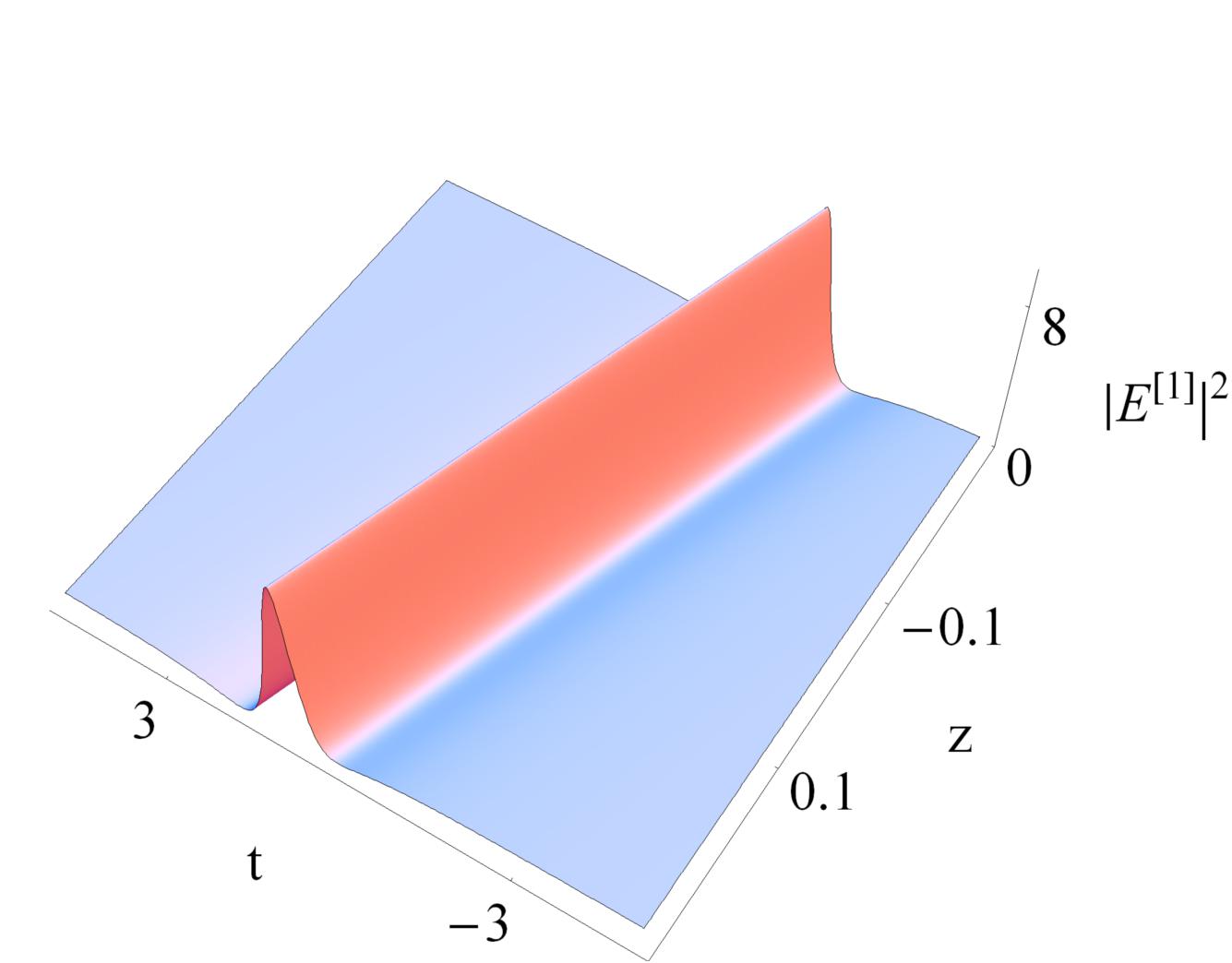}
\quad\quad
\includegraphics[width=100 bp]{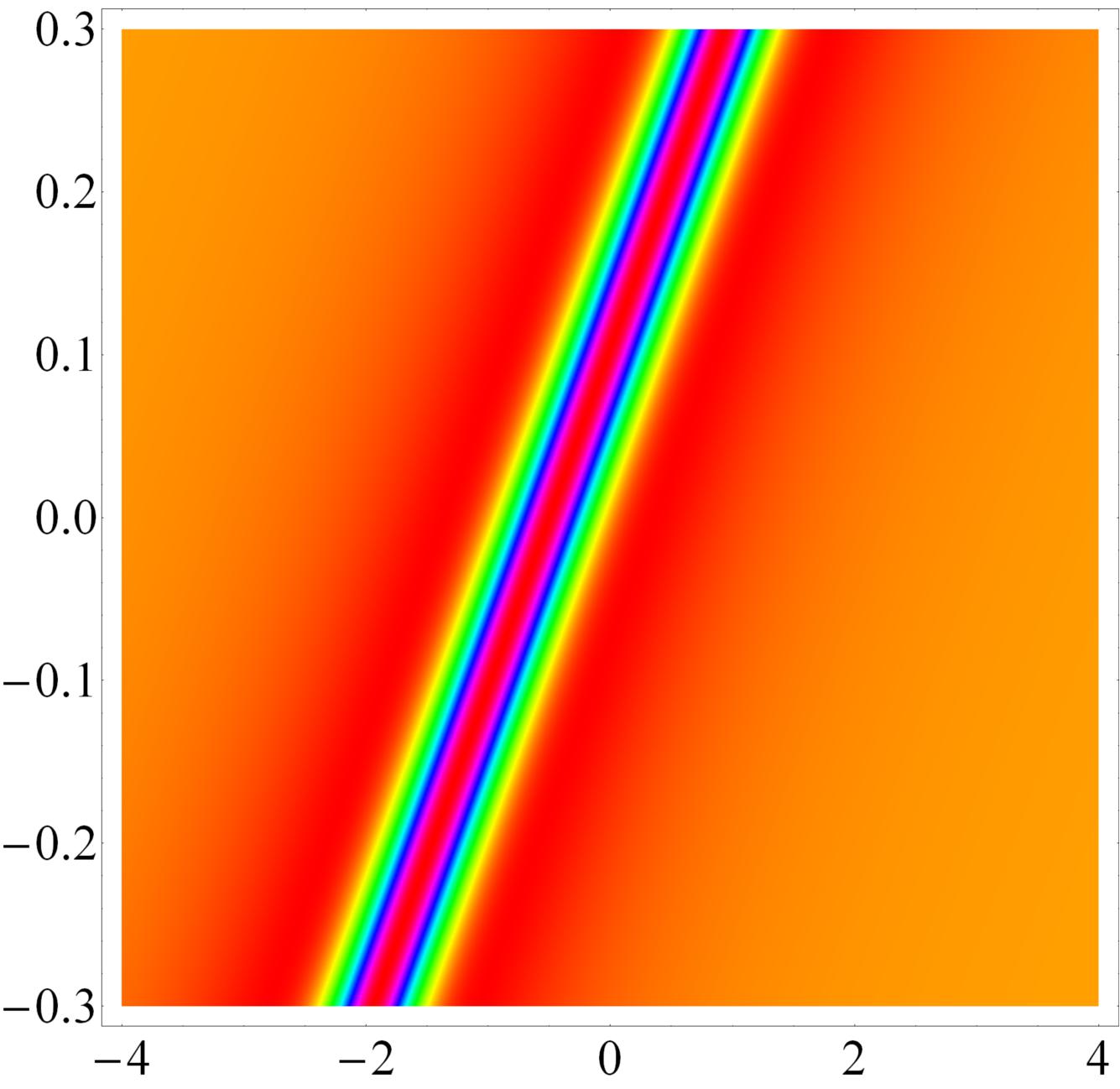}
\includegraphics[height=100 bp]{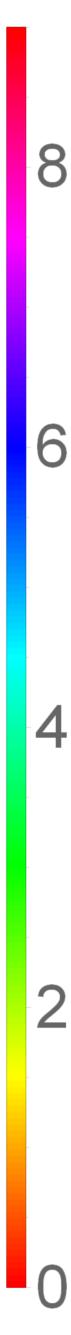}
\quad\quad\quad\quad
\includegraphics[width=150 bp]{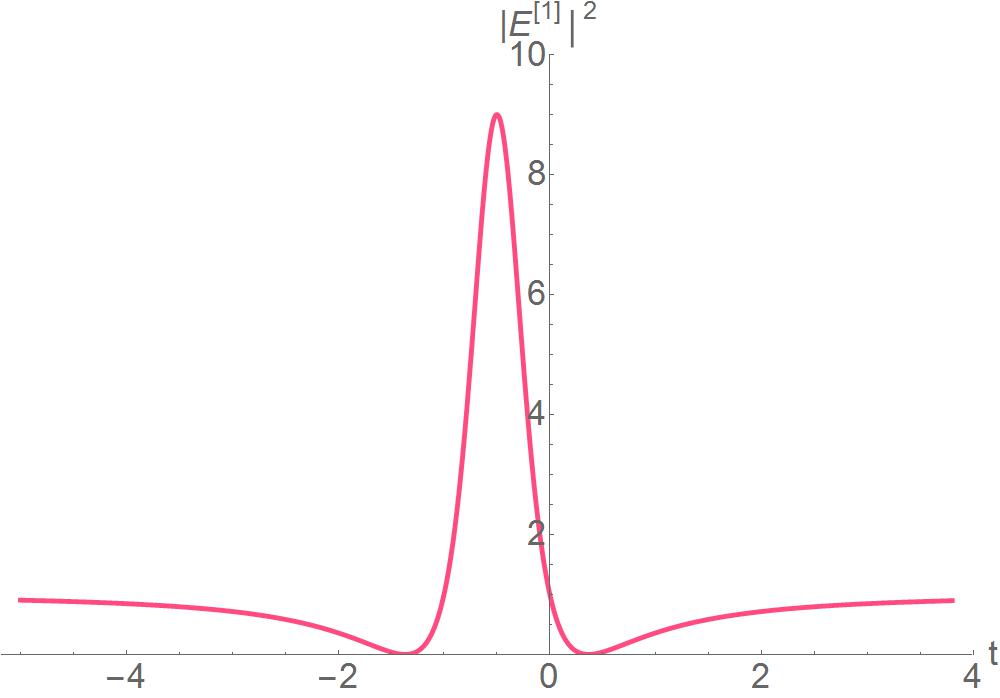}
{\footnotesize  \hspace{4.5cm} Fig.~10(a) \hspace{4cm} Fig.~10(b) \hspace{4.5cm}Fig.~10(c)}
\caption{\footnotesize An asymmetric RW transformed into a W-shaped soliton  with $d=1,b=0,k_{1}=-k_{2}=1,\omega=\frac{1}{2}$ and $\lambda_{0}=-\frac{1}{2}\,b+i\,d$. (b) is the contour plot of (a).\,(c) is the cross-sectional view of (a) at $z=0$. }
\end{figure*}

\end{document}